\documentclass[amsmath,amssymb,aps,pre,floatfix,twocolumn]{revtex4-1}

\usepackage{graphicx}
\usepackage{dcolumn}
\usepackage{bm}
\usepackage{subfigure}  
\usepackage[title]{appendix} 

\newcommand{\LL}{\textsf{L}}

\usepackage{exscale}
\usepackage{multirow} 
\usepackage{booktabs} 

\usepackage{xcolor}
\usepackage[plainpages=false,pdfpagelabels,colorlinks=true,linkcolor=red,urlcolor=blue,citecolor=blue,pdftitle={},pdfauthor={},pdfdisplaydoctitle=true,pdfduplex=DuplexFlipLongEdge]{hyperref}

\definecolor{darkred}{rgb}{0.90,0.2,0.2}

\begin{document}

\preprint{APS/123-QED}

\title{Numerical linked-cluster expansions for two-dimensional spin models\\ with continuous disorder distributions}
\author{Mahmoud Abdelshafy}

\affiliation{Department of Physics, The Pennsylvania State University, University Park, Pennsylvania 16802, USA}
\author{Marcos Rigol}
\affiliation{Department of Physics, The Pennsylvania State University, University Park, Pennsylvania 16802, USA}

\date{\today}

\begin{abstract}
We show that numerical linked cluster expansions (NLCEs) based on sufficiently large building blocks allow one to obtain accurate low-temperature results for the thermodynamic properties of spin lattice models with continuous disorder distributions. Specifically, we show that such results can be obtained computing the disorder averages in the NLCE clusters before calculating their weights. We provide a proof of concept using three different NLCEs based on {\LL}, square, and rectangle building blocks.  We consider both classical (Ising) and quantum (Heisenberg) spin-$\frac{1}{2}$ models and show that convergence can be achieved down to temperatures that are up to two orders of magnitude lower than the relevant energy scale in the model. Additionally, we provide evidence that in one dimension one can obtain accurate results for observables such as the energy down to their ground-state values. 
\end{abstract}

\maketitle

\section{Introduction}
Disorder, resulting from lattice impurities, distortions, or vacancies, can alter the properties of materials in a drastic fashion. For example, noninteracting electrons in the presence of disorder can exhibit Anderson localization~\cite{Anderson}. In spin models, the focus of our work, quenched disorder---affecting the spin exchange interactions---can lead to frustration and spin glasses. Frustration can preclude spin ordering as the temperature is decreased and, below a critical temperature whose value depends on the specific model, a spin glass may form~\cite{Edwards_1975, Binder_spinglass, Weissman_spinglass}. 

The effect of disorder on the thermodynamic properties of quantum spin models remains a challenging topic of current research. Because of frustration, computational approaches such as quantum Monte Carlo techniques suffer from the sign problem, which prevents accessing low-temperature regimes in large system sizes~\cite{loh_90, henelius_00, troyer_05}. Because of the exponential growth of the Hilbert space in quantum systems, exact diagonalization calculations are limited to small system sizes and, due to finite-size effects, it is difficult to make predictions for the behavior of thermodynamic quantities in the thermodynamic limit. Those general limitations for quantum systems with frustration are compounded with the fact that, whenever disorder is present, one needs to carry out calculations for many realizations of disorder and then average over them.

In this work, we show that numerical linked-cluster expansions (NLCEs) can be used to obtain accurate low-temperature results for the thermodynamic properties of classical and quantum spin models with continuous disorder distributions. Previous studies have already shown that NLCEs can be used to obtain accurate results for bimodal~\cite{Tang_dynamicdis, Tang_BiDisorder} and multimodal~\cite{Mulanix_mltidis, park_khatami_21} disorder distributions and that increasing the number of modes in properly selected multimodal disorder distributions can be used to approximate the results for continuous disorder distributions~\cite{Mulanix_mltidis, park_khatami_21}. Our goal here is to show that NLCEs based on large building blocks, such as {\LL}s, squares, and rectangles can be used to carry out direct samplings of continuous disorder distributions to produce accurate results for thermodynamic properties at low temperatures.

NLCEs were originally introduced to study the thermodynamic properties of translationally invariant lattice models in the thermodynamic limit~\cite{rigol_bryant_06, rigol_bryant_07a, rigol_bryant_07b}. They have been broadly used to study clean spin and fermion models since then, see, e.g., Ref.~\cite{Abdelshafy} and references therein. As pointed out in Ref.~\cite{Tang_dynamicdis}, the same NLCEs that are used for translationally invariant systems can be used for bimodal (or multimodal) disorder distributions because the equations defining the linked cluster expansion are linear and averaging over all possible disorder realizations (which are exponentially many but finite for any finite cluster) restores translational symmetry.  

The same applies, in principle, to continuous disorder distributions. However, for continuous distributions, it is impossible to carry out the exact disorder averages except for small clusters. Averages over finite numbers of disorder realizations carry statistical errors that result in a divergence of the NLCEs for the commonly used bond and site expansions introduced in Refs.~\cite{rigol_bryant_06, rigol_bryant_07a}. Divergences occur because computing the weights of large clusters in such expansions involves subtracting weights of exponentially many smaller subclusters, whose statistical errors add up. An alternative way to proceed is to carry out subtractions directly for any given disorder realization on any given cluster (there are no statistical errors in that case) and then average over disorder realizations for that cluster~\cite{devakul_15, Hazzard_dis}. This is computationally very demanding and has yet to be successfully implemented in the context of thermodynamic properties of quantum models with continuous disorder distributions at finite temperature (see Ref.~\cite{Pardini_19} for ground-state calculations).

Here we show that one can overcome the challenges generated by the statistical errors, once again, a consequence of the finite number of disorder realizations that can be computed in models with continuous disorder distributions, using NLCEs with large building blocks. In such NLCEs, the number of clusters grows slowly enough that by solving exactly the smallest clusters and controlling the statistical errors of the clusters that cannot be solved exactly, one can carry out calculations that converge at low temperatures both for classical and quantum spin models. In some cases we find convergence all the way to the ground state. We consider three different NLCEs based on building blocks larger than bonds and sites: a restricted version of the {\LL} expansion introduced recently in Ref.~\cite{Abdelshafy}, the NLCE based on corner-sharing squares introduced in Ref.~\cite{rigol_bryant_07a}, and the rectangle NLCE introduced in Refs.~\cite{kallin_13, Seing_Rect, Hazzard_dis}.

The presentation is organized as follows. The spin-$\frac{1}{2}$ Ising and Heisenberg Hamiltonians studied in this work are introduced in Sec.~\ref{sec:Ham}. A brief review of NLCEs, the resummation algorithms used, and the observables that we calculate is provided in Sec.~\ref{sec:methods}. To build up to our work in the two-dimensional (2D) square lattice, in Sec.~\ref{sec:1D} we use the linked cluster theorem to find a closed form expression for the thermodynamic properties of the 1D Ising model with an arbitrary disorder distribution, as well as NLCEs to numerically study the 1D Heisenberg model in the presence of a uniform disorder distribution. The specific expansions used here in 2D--the restricted {\LL}, the square, and the rectangle expansions--are discussed in Sec.~\ref{sec:lnlce}. In Sec.~\ref{sec:benchmark}, we compare results obtained using those three expansions against site-expansion results reported in Ref.~\cite{Tang_BiDisorder} for the Ising and Heisenberg models with a bimodal disorder distribution. The results for the Ising and Heisenberg models with uniform disorder distributions are reported in Sec.~\ref{sec:results}. We conclude with a summary and discussion of our results in Sec.~\ref{sec:summary}. 

\section{Model Hamiltonians}\label{sec:Ham}

We focus on two spin-$\frac{1}{2}$ Hamiltonians in the thermodynamic limit. The first one is the (classical) Ising model
\begin{equation}
\hat{H}=\sum_{\langle {\bf i},{\bf j}\rangle} J_{\bf ij} \, \hat{S}^{z}_{\bf i}\hat{S}^{z}_{\bf j}\, ,
\label{eq:HIsing}
\end{equation}
where $\hat{S}^{z}_{\bf i}$ is the $z$ component of the spin-$\frac{1}{2}$ operator at site ${\bf i}$ and $\langle {\bf i},{\bf j}\rangle$ denotes pairs of nearest-neighbors sites. Note that the interaction strength $J_{\bf ij}$ depends on the pair of sites $\langle {\bf i},{\bf j}\rangle$. We draw $J_{\bf ij}$ from different discrete and continuous disorder distributions, as specified later. 

We also study the (quantum) Heisenberg model
\begin{equation}
\hat{H}=\sum_{\langle {\bf i},{\bf j}\rangle} J_{\bf ij} \, \hat{\vec {S}}_{\bf i} \cdot \hat{\vec {S}}_{\bf j} ,
\label{eq:HHeis}
\end{equation}
where $\hat{\vec {S}}_{\bf i}$ is now the full spin-$\frac{1}{2}$ operator at site $\bf i$. It follows from the Mermin–Wagner theorem that the Heisenberg model, which has $SU(2)$ symmetry, can only develop long-range order at zero temperature. As for the Ising model, we draw $J_{\bf ij}$ from different discrete and continuous disorder distributions that are specified later. 

\section{A short summary of NLCEs}\label{sec:methods}

NLCEs allow one to calculate finite-temperature properties of extensive observables for translationally invariant lattice models in the thermodynamic limit. For an extensive observable $\mathcal{O}$, its corresponding intensive counterpart per lattice site $O\equiv\mathcal{O}/N$ can be computed using the linked cluster theorem, namely using the following sum over all the connected clusters that can be embedded on the lattice:
\begin{equation}\label{eq:nlce}
O=\sum_{c} L(c)\times W_{\mathcal{O}}(c),
\end{equation}
where $L(c)$ counts the number of ways per site that cluster $c$ can be embedded on the lattice, and $W_{\mathcal{O}}(c)$ is the weight of observable $\mathcal{O}$ in cluster $c$. The weights are calculated recursively via
\begin{equation}\label{eq:weight}
W_{\mathcal{O}}(c)=\mathcal{O}(c)- \sum_{s\subset c} W_{\mathcal{O}}(s),
\end{equation}
with $W_{\mathcal{O}}(c)=\mathcal{O}(c)$ for the smallest cluster.

In NLCEs, one truncates the sum in Eq.~\eqref{eq:nlce} to include only the clusters that can be solved exactly numerically. Convergence at any given temperature $T$ is achieved when the results of successive orders, labelled by the largest clusters considered, agree with each other. For unordered phases, the NLCE results have been shown to approach the thermodynamic limit results exponentially fast in the NLCE order~\cite{iyer_15}. 

Because of the lack of translational invariance in models with disorder, one may think that completely independent NLCE calculations need to be carried out for each disorder realization, so that weights can be properly subtracted. However, as noted in Ref.~\cite{Tang_BiDisorder} in the context of bimodal disorder distributions, averaging over all possible disorder realizations restores translational invariance. This, together with the linear character of the NLCE Eqs.~\eqref{eq:nlce} and~\eqref{eq:weight}, allows one to use the exact same expansion as for translationally invariant models. We use that approach here. Namely, we use Eqs.~\eqref{eq:nlce} and~\eqref{eq:weight} after replacing the expectation values of the observables in each cluster $c$ by their disorder averages $\overline{\mathcal{O}(c)}$. For discrete disorder distributions, such as bimodal disorder, the averages can be computed exactly~\cite{Tang_BiDisorder}. For continuous disorder distributions, we set a maximum value of the normalized standard deviation for all clusters of any given size and explore the effect that changing such a maximum (which in general depends on the cluster size) has on the NLCE results. 

Our calculations are carried out in thermal equilibrium in the grand canonical ensemble at zero chemical potential, so that the many-body density matrix has the form
\begin{equation}\label{eq:rho}
    \hat{\rho} = \frac{1}{Z} \exp \left( -\frac{\hat{H}}{k_BT} \right),\ \text{with} \
    Z=\text{Tr} \left[ \exp \left( -\frac{\hat{H}}{k_BT}\right) \right]\,,
\end{equation}
where $\hat{H}$ is the model Hamiltonian, $k_B$ is the Boltzmann constant (we set $k_B=1$), and $T$ is the temperature (which has units of energy in our convention). We compute three thermodynamic quantities, the energy $E$, the entropy $S$, and the specific heat $C_v$, all per site. 

To gauge the convergence of the direct sums in Eq.~\eqref{eq:nlce}, we calculate the normalized difference for each order $l$ with respect to the highest-order $l_\text{max}$ accessible to us
\begin{equation}\label{rel_error_TH}
\Delta_{l}(O)=\left|\frac{O_{l_\text{max}}-O_{l}}{O_{l_\text{max}}}\right|.
\end{equation}

In order to obtain results at temperatures lower than those at which the direct sums converge, we use resummation techniques. Specifically, we use the following two resummation techniques~\cite{rigol_bryant_07a}: \\
(i) Wynn's ($\epsilon$) algorithm, in which given the original sequence $\{O_{l}\}$,
\begin{eqnarray}
&&\epsilon_l^{(k)}=\epsilon_{l+1}^{(k-2)}+\frac{1}{\epsilon_{l+1}^{(k-1)}-\epsilon_l^{(k-1)}},\\
&&\text{with\quad}\epsilon_l^{(-1)}=0, \quad \epsilon_l^{(0)}=O_l,\nonumber 
\end{eqnarray}
where $k$ denotes the number of Wynn resummation ``cycles.'' Only even entries $\epsilon_{l}^{(2k')}$ (with $k'$ an integer) are expected to converge to the thermodynamic limit result. We note that the new sequence generated after two cycles has two fewer terms. The estimate for an observable after $2k'$ cycles is given by
\begin{equation}
\text{Wynn}_{k'}(O)=\epsilon_{l_\text{max}-2k'}^{2k'}.
\end{equation} 
where we call $k'$ the Wynn resummation ``order''.\\
(ii) The Euler algorithm, which can accelerate the convergence of alternating series. In this algorithm, see Ref.~\cite{Tang_2013}, the only free parameter is the number of terms ``$k$'' for which the direct sum is carried out before the Euler transformation is used. In what follows whenever we report the results of the Euler algorithm, $\text{Euler}_k(O)$, we specify the value of $k$ used.

For further details about NLCEs and their convergence, as well as about the resummation techniques used, we refer readers to the pedagogical introduction in Ref.~\cite{Tang_2013}.

\section{Ising and Heisenberg models in 1D}\label{sec:1D}

In this section, we study the thermodynamic properties of the Ising and Heisenberg models with continuous disorder distributions in 1D. We note that, in 1D, pairs of nearest-neighbors sites $\langle {\bf i},{\bf j}\rangle\equiv i, i+1$, i.e., we can parametrize the bonds $J_{\bf ij}$ with one index and write $J_i$. 

\subsection{Ising model}

The (classical) 1D Ising model with a continuous disorder distribution is exactly solvable for any probability distribution function (PDF) $P(J_{i})$~\cite{Kowalski_1975}. 

\subsubsection{Exact solution}
For an open chain with $N$ sites, using the traditional transfer matrix method it is straightforward to calculate the partition function $Z_{N}=2\prod_{i=1}^{N-1}2 \cosh(\beta J_{i})$, where $\beta = 1/T$ (we set $k_{B} = 1$). One can therefore calculate the intensive quantity $\ln(Z_N)/N$ and average over the PDF in the limit $N \rightarrow \infty$,
\begin{eqnarray}
    \lim_{N\rightarrow \infty} \! \left[ \frac{\overline {\ln(Z_{N})}}{N} \right] \! && \! = \! \lim_{N\rightarrow \infty } \! \frac{1}{N} \! \sum_{i=1}^{N-1} \! \int \! \ln[2 \cosh(\beta J_{i})]P(J_i)dJ_i \nonumber \\
    && + \lim_{N\rightarrow \infty } \frac{\ln2}{N} .
\end{eqnarray}
We therefore have $N-1$ identical integrals so, in the thermodynamic limit, the previous equation simplifies to 
\begin{equation}
    \left [\frac{\overline {\ln(Z_{N})}}{N} \right ] = \ln 2 + \int \ln \left[ \cosh(\beta J) \right] P(J) dJ.
    \label{eq:FIS_1D_ex}
\end{equation}
The free energy is $F = -\ln(Z)/\beta$ so, using Eq.~\eqref{eq:FIS_1D_ex}, one can obtain other thermodynamic properties computing derivatives of the free energy.

\subsubsection{Linked cluster theorem solution}

For a single site we only have one configuration, with two possible states, so the partition function is trivially $\ln(Z_{1})=\ln 2$. For two sites, the partition function is $Z_{2}=2(e^{\beta J}+e^{-\beta J})$, where $J$ is a random coupling constant chosen from the PDF $P(J)$. We then average over all possible $J$'s to get the average $\ln(Z_2)$
\begin{eqnarray}
    \overline{\ln(Z_2)} &&=\int \ln[2(e^{\beta J}+e^{-\beta J})]P(J)dJ \nonumber \\ 
    && = \ln2 + \int \ln(e^{\beta J}+e^{-\beta J})P(J)dJ.   
\end{eqnarray}
For the open chain with three sites, there are two different coupling constants, $J_1$ and $J_2$, drawn from identical independent PDFs $P(J)$. The partition function follows
\begin{equation}
    Z_{3} = 2(e^{\beta J_1}+e^{-\beta J_1})(e^{\beta J_2}+e^{-\beta J_2}).
\end{equation}
Taking the average for $\ln(Z_3)$, as we did for the two-site chain, we get
\begin{eqnarray}
    \overline{\ln(Z_3)} &&=\ln 2 + \int \ln(e^{\beta J_1}+e^{-\beta J_1})P(J_1)P(J_2) dJ_{1}dJ_{2} \nonumber \\
    &&+\int \ln(e^{\beta J_2}+e^{-\beta J_2})P(J_1)P(J_2)dJ_{1}dJ_{2} \nonumber \\
    &&= \ln2 + 2\int \ln(e^{\beta J}+e^{-\beta J})P(J)dJ.
\end{eqnarray}
Such a simple result is a consequence of the factorizable nature of the partition function in the bond strengths, which is unique to the Ising model because of the absence of cross terms. This holds true for chains with an arbitrary number of bonds $N-1$. 

We are ready to calculate the weights defined in Eq.~\eqref{eq:weight}
\begin{eqnarray}
    &&\overline{W_{1}}\!=\!\overline{\ln(Z_{1})}\!=\!\ln(Z_1)\!=\!\ln2 \nonumber \\
    &&\overline{W_{2}}\!=\!\overline{\ln(Z_{2})}-2\overline{W_{1}}\!=\! \! \int \! \ln(e^{\beta J}+e^{-\beta J})P(J)dJ-\ln(2) \nonumber \\
    &&\overline{W_{3}}\!=\!\overline{\ln(Z_{3})}-2\overline{W_{2}}-3\overline{W_{1}}\!=\!0.
\end{eqnarray}
Starting with $W_{3}$, due to the factorizable nature of the partition function, the weights for all orders of the NLCE vanish. We therefore get 
\begin{eqnarray}
    \frac{\overline{\ln(Z)}}{N}&=&\sum_{i}W_i =\int \ln(e^{\beta J}+e^{-\beta J})P(J)dJ \nonumber \\
    &=& \int \ln[2\cosh(\beta J)]P(J)dJ ,
\end{eqnarray}
which is the exact solution, see Eq.~\eqref{eq:FIS_1D_ex}. Hence, like for the clean Ising model~\cite{iyer_15}, the linked cluster theorem allows one to find the exact solution for the 1D Ising model with an arbitrary disorder distribution.

\subsection{Heisenberg model}\label{sec:1dheisenberg}

In contrast to the translationally invariant case, the (quantum) 1D Heisenberg model with a continuous disorder distribution is not exactly solvable~\cite{Ma_79, Dasgupta_Ma_80}. We study this model numerically using NLCEs and focus on the zero-mean uniform disorder distribution, with PDF
\begin{equation}
    P(x) = \begin{cases}
 &\frac{1}{2J} \quad \text{for}\quad -J \leq x \leq  J \\ 
 &\ 0\ \quad \text{for} \quad \left | x \right | > J
\end{cases}.
\label{eq:uniform}
\end{equation}

\begin{figure}[!t]
    \includegraphics[width=.98\columnwidth]{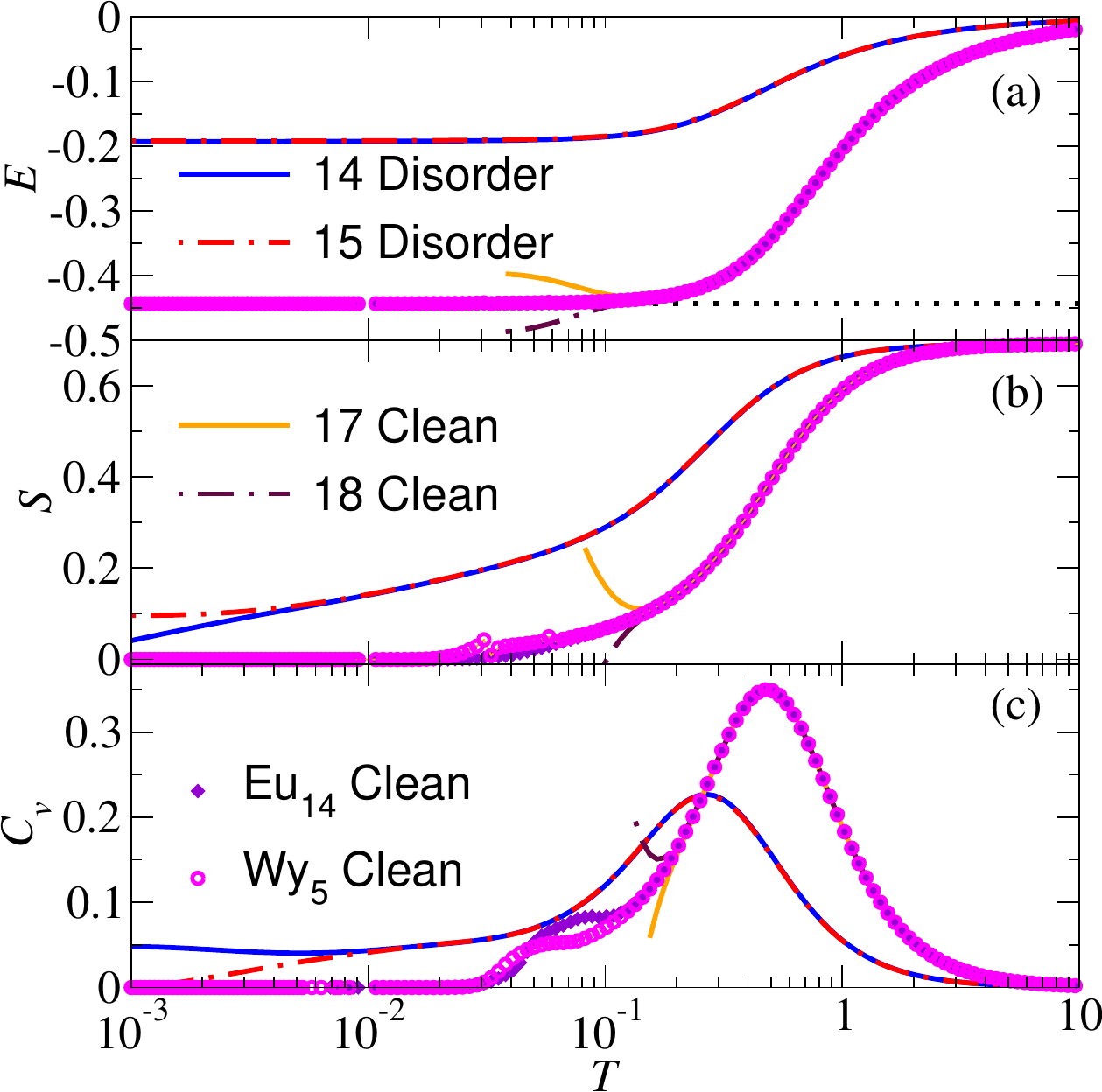}
    \vspace{-0.1cm}
    \caption{Thermodynamic properties of the 1D Heisenberg model with a uniform disorder distribution, and for the clean Heisenberg model. We plot the NLCE results for (a) energy $E$, (b) entropy $S$, and (c) specific heat $C_v$ vs $T$ obtained at orders $l=14$ and 15 ($l=17$ and 18) for the model with disorder (clean model), and Wynn’s and Euler’s resummation results for the clean model. The horizontal dotted line in (a) shows the exact result for the ground-state energy of the clean model, $E_G=1/4-\ln2\approx -0.443$~\cite{Giamarchi_03}.}
    \label{fig:1D_AF}
\end{figure}

In 1D models with only nearest-neighbor couplings, there is one cluster at each order $l$ of the NLCE; an open chain with $l$ sites. When calculating the $l\text{th}$ order NLCE result for an observable, Eq.~\eqref{eq:nlce} simplifies to $O_l=\overline{\mathcal{O}(l)}-\overline{\mathcal{O}(l-1)}$. So one only needs to calculate observables for two consecutively cluster sizes in order to get the NLCE result at any given order. In Fig.~\ref{fig:1D_AF}, we plot NLCE results for the energy $E$, the entropy $S$, and the specific heat $C_v$ vs $T$ for the 1D Heisenberg model with a uniform disorder distribution. We set $J=1$ to be our energy scale and report results for $l=14$ and $l=15$. Those results were obtained carrying out averages over 17.5, 4, and 0.6 million disorder realizations for the chains with 13, 14, and 15 sites, respectively. We also show the corresponding results for the clean model with $J=1$.

For the energy in the presence of disorder [Fig.~\ref{fig:1D_AF}(a)], we find that the results for $l=14$ and $l=15$ agree with each other down to $T=10^{-3}$, at which $E$ has become temperature independent and we essentially obtain the ground-state energy. This is to be contrasted to the results for the clean model, for which the results for $l=17$ and $l=18$ agree with each other down only to $T\approx0.2$. Disorder, which reduces correlations, extends the NLCE convergence to lower temperatures for all the observables considered here. Resummations for the clean model do allow one to reproduce the exact ground-state energy (shown as a horizontal dotted line). For the entropy in the presence of disorder [Fig.~\ref{fig:1D_AF}(b)], the results for $l=14$ and $l=15$ agree with each other down to $T\approx 10^{-2}$, in comparison to $T\approx0.2$ for the clean model. The contrast between the results in the presence and absence of disorder make apparent that the relatively high value of the entropy at $T\approx 10^{-2}$ in the former is a consequence of frustration introduced by the random couplings. In the clean model, the resummation results indicate that the entropy at that temperature is vanishingly small. Like for the entropy, for the specific heat in the presence of disorder [Fig.~\ref{fig:1D_AF}(c)], the results for $l=14$ and $l=15$ agree with each other down to $T\approx 10^{-2}$. NLCEs show that there is a well-resolved peak in the specific heat with a maximum value $C^{\text{max}}_v \approx 0.23$ at $T_m \approx 0.25$. On the other hand, the peak appearing in the clean model has a maximum $C^{\text{max}}_v \approx 0.37$ at $T_m \approx 0.45$. Disorder reduces the height of that peak and moves it towards lower temperatures.

Next, we compare convergence errors [see Eq.~\eqref{rel_error_TH}] to normalized standard deviations
\begin{equation}\label{eq:staterror}
\delta_{c}(\mathcal{O}) = \frac{\sigma_c({\mathcal{O}})}{{\overline{\mathcal{O}(c)}}}\,,
\end{equation}
where
\begin{equation}
\sigma_c({\mathcal{O}})=\sqrt{\overline{\mathcal{O}^2(c)}-\overline{\mathcal{O}(c)}^2},
\end{equation}
with ${\overline{\mathcal{O}(c)}}$ being the disorder average of the observable in cluster $c$, to gain insights on the temperatures at which lack of convergence due to the size of the clusters dominates over statistical errors, and vice versa.

\begin{figure}[!t]
    \includegraphics[width=.98\columnwidth]{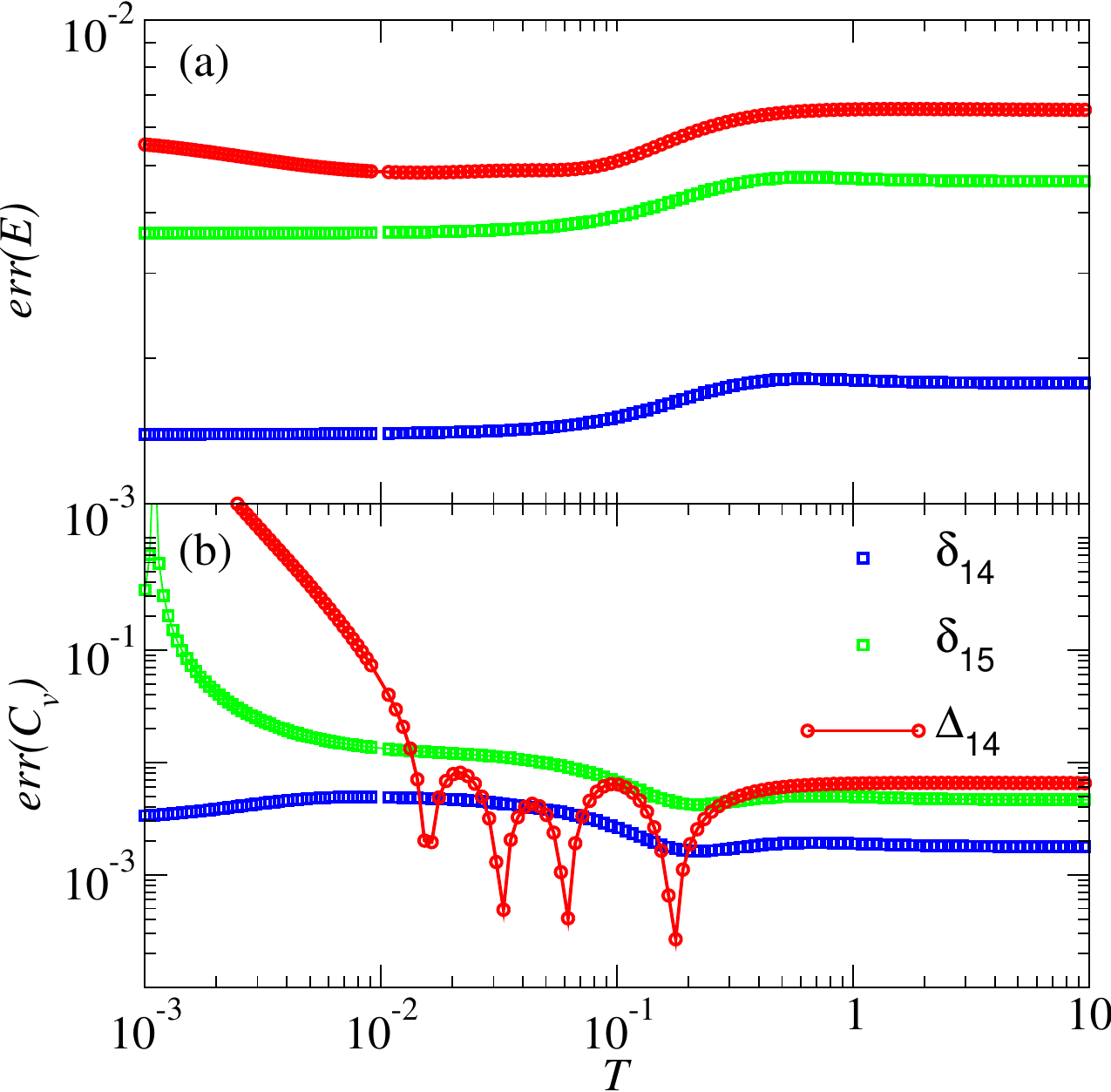}
    \vspace{-0.1cm}
    \caption{Errors in the NLCE calculations of thermodynamic properties of the 1D Heisenberg model with a uniform disorder distribution. Differences $\Delta_{14}(O)$, see Eq.~\eqref{rel_error_TH}, and the normalized standard deviations $\delta_{c}(\mathcal{O})$, see Eq.~\eqref{eq:staterror}, for the chain clusters $c=14$ and 15 with 14 and 15 sites, respectively. (a) Energy $E$ and (b) specific heat $C_v$.}
    \label{fig:1D_AF_err}
\end{figure}

In Fig.~\ref{fig:1D_AF_err}, we plot $\Delta_{14}$ and $\delta_{c}$ for the chain clusters with 14 and 15 sites (the two largest ones considered), both for the energy and the specific heat (the results for the entropy, not shown, are qualitatively similar to those for the specific heat). For the energy [Fig.~\ref{fig:1D_AF_err}(a)], $\Delta_{14}(E)$ is of the order of (slightly larger than) the normalized standard deviations for the two clusters at the temperatures shown. This makes apparent that the statistical errors are the main errors in the NLCE calculations of the energy at those temperatures. For the specific heat, the results in Figs.~\ref{fig:1D_AF_err}(b) show that the statistical errors are the main errors only at temperatures $T\gtrsim 0.01$ [at the temperatures at which the NLCE results for orders 14 and 15 are indistinguishable from each other in Fig.~\ref{fig:1D_AF}(c)]. At very low temperatures $T\lesssim 0.01$, $\Delta_{14}(C_v)$ becomes much larger than the statistical errors, which shows that lack of convergence due to the size of the clusters dominates the error in the NLCE calculations in that regime (as it does in the clean case).

\section{NLCEs in 2D}\label{sec:lnlce}

There are various NLCEs based on different building blocks that have been used in the literature to study square lattice models. Here we focus on three schemes. First, we introduce a restricted {\LL} expansion with a significantly lower number of clusters than the {\LL} expansion introduced in Ref.~\cite{Abdelshafy}. We further use the square expansion introduced in Ref.~\cite{rigol_bryant_07a}, and the rectangle expansion introduced in Refs.~\cite{kallin_13,  Seing_Rect, Hazzard_dis}. For comparison, we also show some results obtained using the site-based expansion introduced in Ref.~\cite{rigol_bryant_07a}.

The main advantage of the former three NLCEs is the low number of clusters that enter in the corresponding expansions up to relatively large cluster sizes. This allows us to control the statistical errors of the average over disorder realizations in each cluster and prevents a significant build up of those errors that would result in divergences of the bare sums. In the following subsections, we briefly introduce those three NLCE schemes.

\subsection{{\LL} expansion}

\begin{figure}[!b]
\centering
\includegraphics[width=0.95\columnwidth]{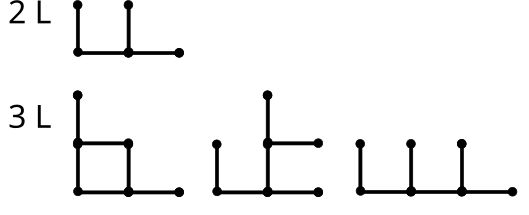}
\vspace{-0.1cm}
\caption{Clusters present in the second (2 {\LL}s) and third (3 {\LL}s) orders of both the strong embedding {\LL} and the restricted {\LL} expansion.}\label{fig:allowed}
\end{figure}

In Ref.~\cite{Abdelshafy}, we developed ``strong'' and ``weak'' embedding versions of an NLCE expansion that uses {\LL}s as building blocks. We showed that the strong-embedding version (with all possible {\LL}s connecting the sites present), which generally involves a smaller number of clusters at each order and hence has a lower computational cost, was preferable as (i) it has similar convergence properties as the weak embedding version in the high-temperature disordered phase, and (ii) it was the only {\LL} expansion that converged when approaching the ground state in ordered phases such as the one in the Ising model. Here, in order to reduce even further the number of clusters of the strong embedding {\LL} expansion, we introduce a {\it restricted} {\LL} expansion. In the restricted expansion, the {\LL}s in the strong embedding NLCE are attached to an existing cluster by sharing the center site, i.e., no {\LL} can share only edge sites in any given cluster.

\begin{table}[!t]
\caption{Total number of clusters (second and third columns) and the number of topologically distinct clusters (fourth and fifth columns) in the restricted (R) and unrestricted (U) strong embedding {\LL} expansions, respectively, versus the number of {\LL}s in the clusters (first column).}
\label{Table_ngraphs_RU} 
\begin{ruledtabular}
\begin{tabular}{r r rr rr}
\multicolumn{2}{l}{No.~{\LL}s}	 & \multicolumn{2}{r}{Total No.~clusters} 	& \multicolumn{2}{r}{No.~top. clusters} \\
\cmidrule(lr){1-2} \cmidrule(lr){3-4} \cmidrule(lr){5-6}
& & R & U & R & U \\
\midrule
\hline
0      &       &1          &1             &1            &1     \\
1      &       &1          &1             &1            &1     \\
2      &       &2          &3             &1            &2     \\   
3      &       &5          &11            &2            &6     \\  
4      &       &13         &41            &4            &18    \\  
5      &       &34         &153           &7            &61    \\ 
6      &       &90         &573           &15           &202   \\ 
7      &       &239        &2162          &30           &700   \\ 
8      &       &636        &8238          &62           &2429  \\ 
9      &       &1695       &31696         &129          &8608  \\
10     &       &4522       &122986        &268          &30734 \\
11     &       &12075      &NA            &562          &NA    \\
12     &       &32265      &NA            &1178         &NA    \\
\end{tabular}
\end{ruledtabular}	
\end{table}

Such a restriction on the clusters allowed in the strong embedding {\LL} expansion results in more compact (larger weight) clusters and reduces the number of clusters significantly. In Fig.~\ref{fig:allowed}, we show the clusters with two and three {\LL}s that are present in the restricted {\LL} expansion. The total number of clusters [which for the {\LL} expansion equals the sum of $L(c)$s in Eq.~\eqref{eq:nlce}] at each order of the restricted {\LL} expansion are shown in the second column of Table~\ref{Table_ngraphs_RU}. Those numbers are to be compared to the total number of clusters in the strong embedding {\LL} expansion~\cite{Abdelshafy} shown in the third column. The fourth and fifth columns in Table~\ref{Table_ngraphs_RU} show the total number of topologically distinct clusters in each expansion, which are the actual clusters that are diagonalized to compute the observables as they are the ones with different Hamiltonians. One can see that there is an exponential reduction of the number of clusters from the unrestricted to the restricted {\LL} expansion as the number of {\LL}s increases. Clusters with the same number of {\LL}s are grouped together and the order $l$ of the expansion is set by the largest number of {\LL}s included in the NLCE sum. 

Among the three main expansions considered in this work, the {\LL} expansion is the one that has the most clusters with any given number of sites. This means that the {\LL} expansion is the one that best explores the square lattice geometry, and we expect it to provide the most accurate results at intermediate and high temperatures.

\subsection{Square expansion}

The square expansion is an expansion based on corner-sharing squares~\cite{rigol_bryant_07a}. In Table~\ref{Table_ngraphs_sq} one can see that, up to six squares (a maximum of 19 sites), it involves a very small number of clusters. Clusters with the same number of squares are grouped together and the order $l$ of the expansion is set by the largest number of squares included in the NLCE sum.

\begin{table}[!t]
\caption{Number of topologically distinct clusters (second column) and the sum of $L(c)$s (third column) in the square expansion versus the number of squares (first column).}
\label{Table_ngraphs_sq} 
\begin{ruledtabular}
\begin{tabular}{r r r}
No.~squares	 &No.~top. clusters	 & Sum of $L(c)$s \\
\hline
0           &1           &1        \\
1           &1           &1/2      \\
2           &1           &1        \\   
3           &2           &3        \\  
4           &5           &19/2     \\  
5           &11          &63/2     \\ 
6           &31          &108      \\ 
\end{tabular}
\end{ruledtabular}	
\end{table}

\subsection{Rectangle expansion}

The rectangle expansion was introduced in Ref.~\cite{kallin_13} to calculate entanglement entropies, and was used in Ref.~\cite{Seing_Rect} to study quench dynamics in clean systems and in Ref.~\cite{Hazzard_dis} to study quench dynamics from inhomogeneous initial states. The rectangle expansion contains clusters that have a rectangular shape. This limits the number of clusters considerably as there are only three possible cluster geometries. For clusters with $N$ sites one can have (i) a chain of $N$ sites; (ii) a rectangle with $N=N_x\times N_y$ sites (for values of $N$ that admit such a decomposition), where $N_x$ and $N_y$ are the numbers of sites in $x$ and $y$, respectively; and (iii) a square with $N=N_x^2$ sites for $N=4, 9,\,\dots$. Squares have $L(c) = 1$, while all other clusters have $L(c) = 2$, making the combinatorics associated with the rectangle expansion trivial. Clusters with the same number of sites are then grouped together and the order $l$ of the expansion is set by the largest number of sites included in the NLCE sum. 

\section{Bimodal disorder distribution}\label{sec:benchmark}

In order to gain an understanding of how the {\LL}, the square, and the rectangle expansions work in the presence of disorder, in this section we compare our results using those expansions for a bimodal disorder distribution to results obtained in Ref.~\cite{Tang_BiDisorder} using the site expansion. For bimodal disorder, each $J_{\bf ij}$ can have values $\pm J$ with equal probability (we set $J=1$). An advantage of such a distribution (e.g., over the continuous ones that we study in the next section) is that one can average over all possible disorder realizations ($2^b$, where $b$ is the number of bonds) in the finite clusters considered in the NLCEs so that there are no errors associated to the sampling.

\begin{figure}[!t]
    \includegraphics[width=.98\columnwidth]{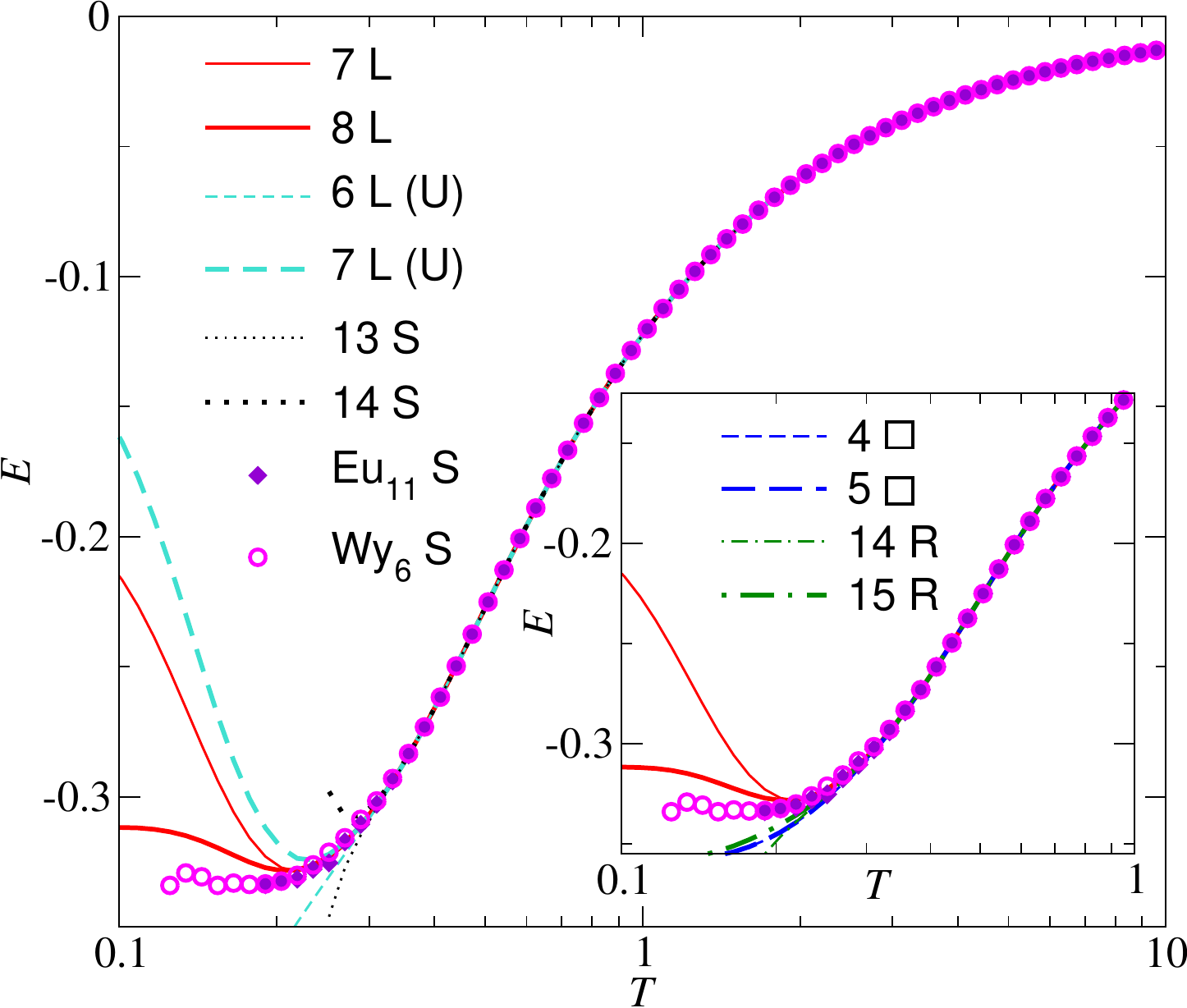}
    \vspace{-0.1cm}
    \caption{Energy per site $E$ vs $T$ for the Ising model with a bimodal disorder distribution. We show results for the restricted {\LL} expansion (\LL) with 7 and 8 {\LL}s, the unrestricted {\LL} expansion [\LL (U)] with 6 and 7 {\LL}s, the site expansion (S) with 13 and 14 sites, and Wynn's and Euler's resummations of the site expansion. All the site expansion results are from Ref.~\cite{Tang_BiDisorder}. Inset: Restricted {\LL} expansion with 7 and 8 {\LL}s, and Wynn's and Euler's resummations of the site expansion (same results and legends as in the main panel) together with the results for the square expansion ($\square$) with 4 and 5 squares, and the rectangle expansion (R) with 14 and 15 sites.}
    \label{fig:IS_bidis}
\end{figure}

In Fig.~\ref{fig:IS_bidis}, we show site expansion results for the energy of the 2D Ising model with a bimodal disorder distribution, along with Euler and Wynn resummation results, reported in Ref.~\cite{Tang_BiDisorder}. The energies from the 13 and 14 orders of the site expansion agree with each other down to $T\approx 0.3$, while the resummation results agree with each other down to $T\approx 0.2$. This means that the direct sums allow one to compute the energies for $T \gtrsim 0.3$, and the resummations allow one to estimate the energies for $0.2\lesssim T \lesssim 0.3$. We also show in Fig.~\ref{fig:IS_bidis} results for the {\LL} expansions. The energies from the (unrestricted) strong embedding {\LL} expansion with up to 6 and 7 {\LL}s, labeled with a ``(U)'' in Fig.~\ref{fig:IS_bidis}, agree with each other down to temperatures slightly lower than those at which the 13 and 14 orders of the site expansion agree with each other. The restricted strong embedding {\LL} expansion results with up to 7 and 8 {\LL}s agree with each other down to slightly lower temperatures than the other two expansions. 

Recall that the number of clusters in the restricted {\LL} expansion grows much more slowly than in the unrestricted one as the number of {\LL}s increases, and this is the reason we can compute one order higher of the former expansion for the results shown in Fig.~\ref{fig:IS_bidis}. Given the excellent convergence properties of the restricted {\LL} expansion, along with the fact that its smaller number of clusters per order will allow us to reduce the effect of statistical errors in the subgraph subtractions later when we study continuous disorder distributions, we focus on that {\LL} expansion in what follows. We will refer to the restricted {\LL} expansion as the {\LL} expansion in the rest of this paper. In the inset in Fig.~\ref{fig:IS_bidis} we compare the Euler and Wynn resummation results for the site expansion with results for the bare sums obtained using the {\LL}, the square, and the rectangle expansions. They all agree with each other down to temperatures that are slightly higher than $0.2$, with the {\LL} expansion results agreeing with the resummation ones at lower temperatures than the square and rectangle expansions. Having the independent results from the {\LL}, the square, and the rectangle expansions will be useful in the rest of this work to gauge convergence for different models.

\begin{figure}[!t]
\includegraphics[width=.97\columnwidth]{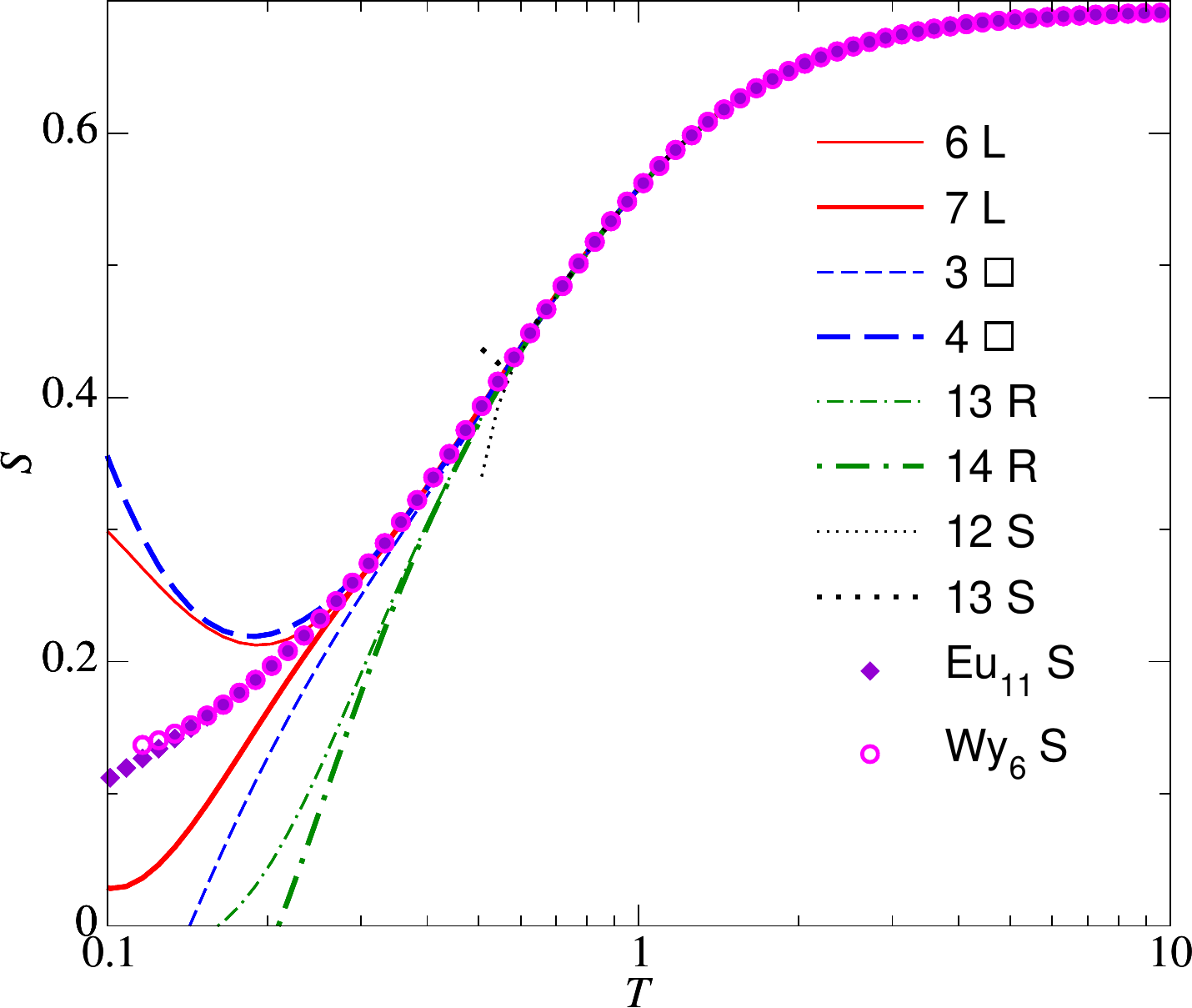}
    \vspace{-0.1cm}
\caption{Entropy per site $S$ vs $T$ for the Heisenberg model with a bimodal disorder distribution. We show results for the restricted {\LL} expansion (\LL) with 6 and 7 {\LL}s, the square expansion ($\square$) with 3 and 4 squares, and the rectangle expansion (R) with 13 and 14 sites, the site expansion (S) with 12 and 13 sites, and Wynn's and Euler's resummations of the site expansion. All the site expansion results are from Ref.~\cite{Tang_BiDisorder}.}\label{fig:AF_S_bidis}
\end{figure}

In Fig.~\ref{fig:AF_S_bidis}, we show site expansion results for the entropy of the 2D Heisenberg model with a bimodal disorder distribution, along with Euler and Wynn resummation results, reported in Ref.~\cite{Tang_BiDisorder}. The entropies from the 12 and 13 orders of the site expansion agree with each other down to $T\approx 0.6$, while the resummation results agree with each other down to $T\lesssim 0.2$. These results make apparent that resummation techniques can provide accurate estimates of thermodynamic quantities at significantly lower temperatures ($\sim$3 times lower in this case) than the direct sums. Remarkably, our results for the last two orders of the {\LL} and the square expansions agree with each other and with the resummation results for the site expansion down to $T\approx 0.3$. Such an agreement makes apparent the effectiveness of NLCE expansions based on {\LL}s and squares in providing converged results at temperatures that are significantly lower than those at which the direct sums for the site expansion converge. The results for the last two orders of the rectangle expansion are close to each other down to $T\approx 0.3$, but they depart from those of the other expansions at temperatures below $T\approx 0.5$. This is an indication that, below $T\approx 0.5$, the rectangle expansion results for the entropy converge slowly with increasing the order of the expansion. 

The results in Fig.~\ref{fig:AF_S_bidis} highlight the importance of using different NLCE schemes together with resummation techniques to gauge convergence. In the context of the expansions used in this work, Fig.~\ref{fig:AF_S_bidis} makes apparent that we need to be especially careful with the rectangle expansion results as they may appear converged at temperatures that they are not. The departure of the rectangle expansion results from those of the {\LL} and square expansions is likely a consequence of the fact that, at the orders considered in the rectangle expansion, chain and ladder clusters are significantly more abundant than square and close to square ones, i.e., there is a ``bias'' towards quasi-1D shaped clusters.

\section{Continuous disorder distribution}\label{sec:results}

Next, we study the thermodynamic properties of the 2D Ising and Heisenberg models with continuous disorder distributions. Our focus is on the case in which the distribution of disorder is uniform, as defined in Eq.~\eqref{eq:uniform}. In contrast to the case of bimodal disorder considered in the previous section, for a continuous disorder distribution it is not possible to compute the exact disorder averages for all the clusters used in any given NLCE. Hence, central to our discussions in what follows will be how to properly deal with the statistical errors generated by the finite number of disorder realizations sampled to compute the disorder averages. 

\subsection{Ising model}

\begin{figure}[!t]
    \includegraphics[width=.98\columnwidth]{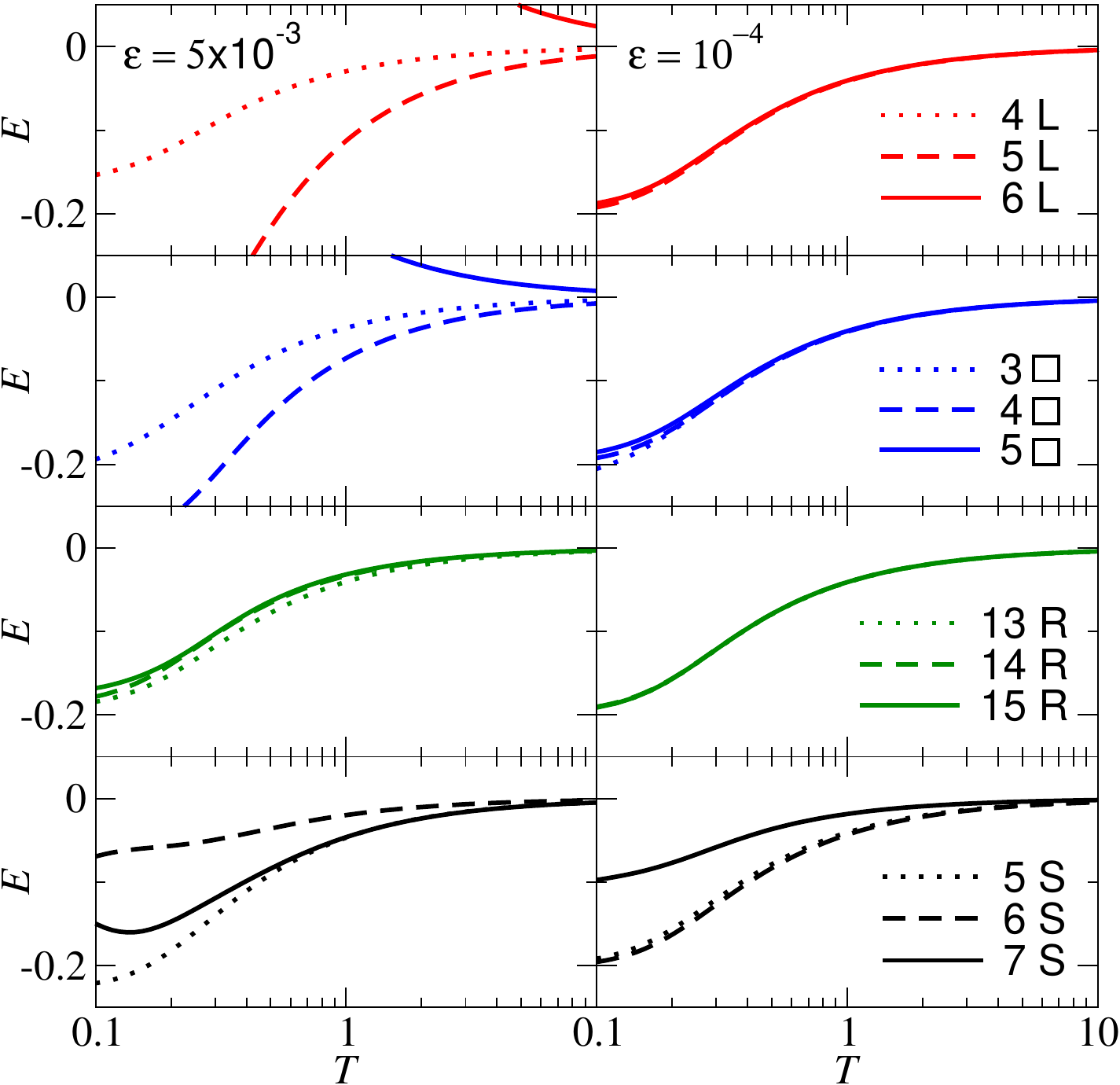}
    \vspace{-0.1cm}
    \caption{Energy per site vs $T$ for the square lattice Ising model with a uniform disorder distribution. The first row shows results for the {\LL} (\LL) expansion, with 4, 5, and 6 {\LL}s. The second row shows results for the square ($\square$) expansion with 3, 4, and 5 squares. The third row shows results for the rectangle (R) expansion with 13, 14, and 15 sites. The fourth row shows results for the site (S) expansion with 5, 6, and 7 sites. The left column shows results for $\varepsilon = 5\times 10^{-3}$, and the right column shows results for $\varepsilon = 10^{-4}$.}
    \label{fig:IS_err_contdis}
\end{figure}

Let us first consider the Ising model with the uniform disorder distribution in Eq.~\eqref{eq:uniform}. To show the effect that decreasing $\delta_{c}$ [see Eq.~\eqref{eq:staterror}] has in our NLCE calculations, in Fig.~\ref{fig:IS_err_contdis} we plot NLCE results for the energy obtained using the {\LL}, square, rectangle, and site expansions (from top to bottom, respectively), when $\delta_{c}(E)\lesssim \varepsilon$ at $T\sim 1$ in all clusters in the expansion. We note that as the size of the clusters increases, because of self-averaging, to achieve the same value of $\delta_{c}(E)$ we need to consider smaller numbers of disorder realizations.

The left (right) panels in Fig.~\ref{fig:IS_err_contdis} show results when $\varepsilon=5\times10^{-3}$ ($\varepsilon=10^{-4}$). For $\varepsilon=5\times10^{-3}$ (left panels in Fig.~\ref{fig:IS_err_contdis}), see Tables~\ref{Table_iterations_all} and~\ref{Table_iterations_Rect} for the number of disorder realizations used, the results for different orders of the expansions differ at temperatures $T>1$, which are sufficiently high for convergence to be achieved given the cluster sizes considered in all the expansions shown, i.e., the lack of convergence observed at those temperatures is a consequence purely of the statistical errors introduced in the averages over finite numbers of disorder realizations. We find the rectangle expansion to be the least affected by those statistical errors for clusters with up to $\sim$15 sites. This is a result of the simple structure of the subgraph subtraction for this expansion, e.g., as for the 1D expansion considered in Sec.~\ref{sec:1dheisenberg}, there is no accumulation of errors for the chain clusters involved in the rectangle expansion. On the other hand, the site expansion results are strongly affected by the statistical errors even for clusters that have about one-half the number of sites of those in the other expansions.

For $\varepsilon=10^{-4}$ (right panels in Fig.~\ref{fig:IS_err_contdis}), see Tables~\ref{Table_iterations_all} and~\ref{Table_iterations_Rect} for the number of disorder realizations used, the results for the {\LL}, the square, and the rectangle expansions agree at temperatures $T>1$ and are very close to each other at temperatures $0.1\leq T\leq 1$. The site-expansion results, on the other hand, still do not agree with each other at temperatures $T>1$. Since statistical errors of the order of $10^{-4}$ require averages over millions of disorder realizations (see Tables~\ref{Table_iterations_all} and~\ref{Table_iterations_Rect}), the lack of convergence of the (low) 7th order of the site expansion makes apparent that such an expansion is not suitable to study models with continuous disorder distributions by averaging over finite numbers of disorder realizations.

\begin{table}[!t]
\caption{Number of disorder realizations used to obtain the results in Fig.~\ref{fig:IS_err_contdis} for the {\LL}, square ($\square$), and site expansions. The numbers, for $\varepsilon=5\times10^{-3}$ ($\varepsilon=10^{-4}$, in parenthesis), are shown in units of $10^{3}$ ($10^{6}$).}
\label{Table_iterations_all} 
\begin{ruledtabular}
\begin{tabular}{r r r r}
Order	    &{\LL}	     & $\square$   & Site \\
\hline
0           &Exact       &Exact        & NA\\
1           &$30$($100$) &$15$($120$)  &Exact\\
2           &$20$($80$)  &$7.5$($100$) &$50$($30$)\\   
3           &$7.5$($55$) &$3.75$($40$) &$40$($30$)\\  
4           &$7.5$($28$) &$3$($5.5$)   &$30$($24$)\\  
5           &$7$($18$)   &$2.5$($1.7$) &$20$($18$)\\ 
6           &$4$($5.5$)  &NA           &$10$($15$)\\ 
7           &NA          &NA           &$4.5$($8$)\\ 
\end{tabular}
\end{ruledtabular}	
\end{table}

\begin{table}[!b]
\caption{Number of disorder realizations used to obtain the results in Fig.~\ref{fig:IS_err_contdis} for the rectangle expansion. The numbers for $\varepsilon=5\times10^{-3}$ and $\varepsilon=10^{-4}$ are shown in the second and third columns, respectively.}
\label{Table_iterations_Rect} 
\begin{ruledtabular}
\begin{tabular}{r r r}
Order	    &$\varepsilon=5\times10^{-3}$     &$\varepsilon=10^{-4}$  \\
\hline
1           &Exact                            &Exact                  \\
2-6         &$10^5$                           &$6\times10^7$          \\
7           &$10^5$                           &$5.5\times10^7$        \\
8           &$10^5$                           &$4\times10^7$          \\
9           &$10^5$                           &$2.8\times10^7$        \\
10          &$10^5$                           &$2\times10^7$          \\
11          &$5\times10^4$                    &$1.8\times10^7$        \\  
12          &$10^4$                           &$10^7$                 \\  
13          &$3\times10^3$                    &$5.5\times10^6$        \\ 
14          &$2.5\times10^3$                  &$4\times10^6$          \\ 
15          &$2\times10^3$                    &$1.7\times10^6$        \\ 
\end{tabular}
\end{ruledtabular}	
\end{table}

In order to improve convergence by reducing the effect of statistical errors even further, we note that the number of subclusters of any given cluster that belongs to order $l$ increases exponentially with $l$, with most of the smallest clusters appearing in the larger clusters. Since the number of clusters also grows exponentially with $l$, the statistical errors of the smallest clusters compound rapidly as the order of the expansion increases. Hence, it is essential to reduce the statistical errors in the smallest clusters as much as possible. To achieve this, in this work, we compute the exact disorder averages for all clusters with up to five sites. Namely, for such clusters, we compute all observables symbolically and then calculate the exact disorder averages by integrating over the continuous disorder distribution. This means that, in what follows, the disorder averages are computed exactly for clusters with one and two {\LL}s in the {\LL} expansion, for one square in the square expansion, and for chains with one through five sites and the square with four sites in the rectangle expansion. For higher orders of these expansions, the number of disorder realizations used is about 10 times the ones reported between parentheses in Table~\ref{Table_iterations_all} and in the right column of Table~\ref{Table_iterations_Rect}.

\begin{figure}[!t]
    \includegraphics[width=.98\columnwidth]{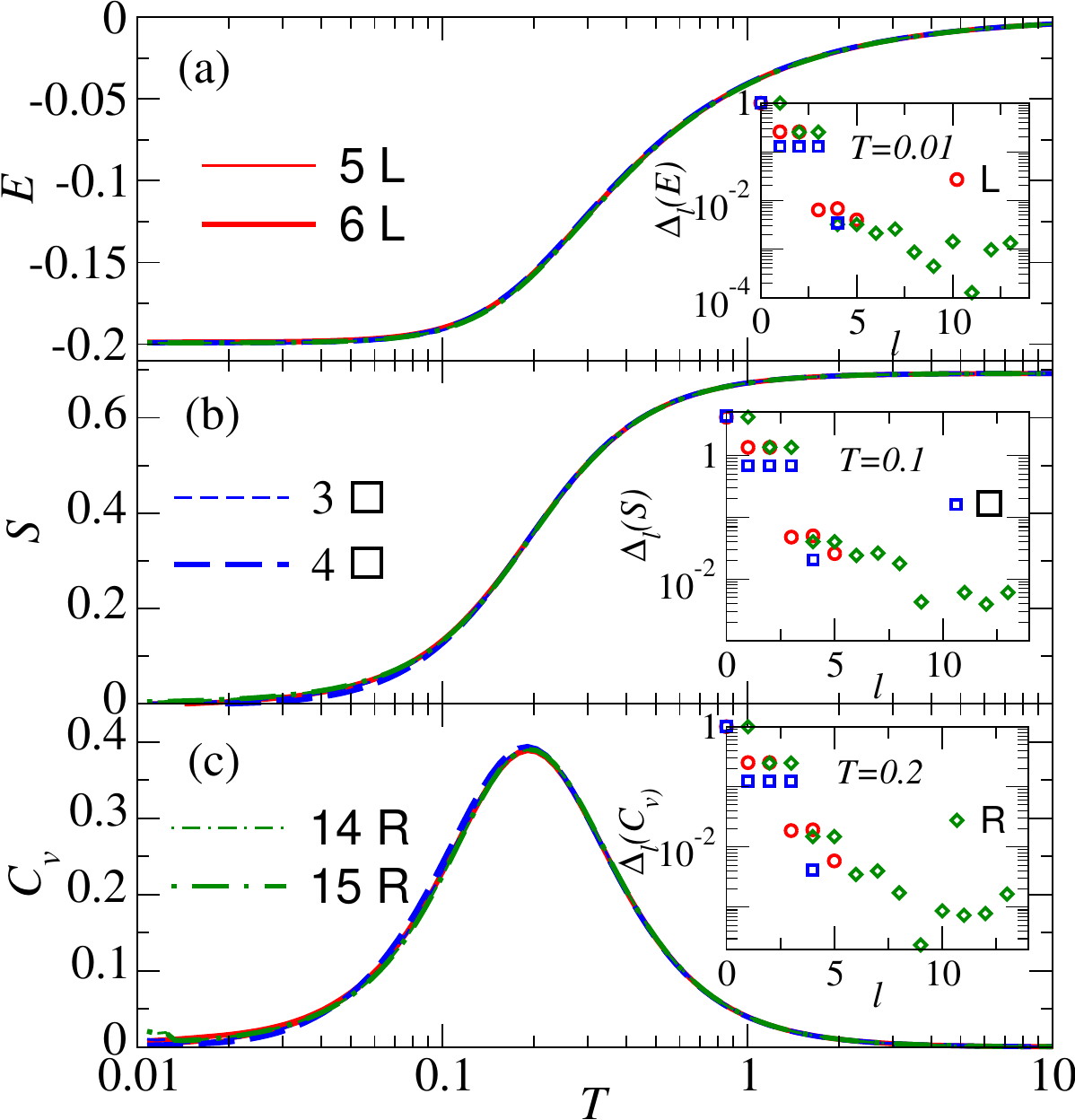}
    \vspace{-0.1cm}
    \caption{Thermodynamic properties of the square lattice Ising model with a uniform disorder distribution. (a) Energy $E$, (b) entropy $S$, and (c) specific heat $C_v$ per site vs $T$ obtained using the {\LL} (\LL), the square ($\square$), and the rectangle (R) expansions. We report results for the highest two orders computed of each NLCE scheme. Insets: Normalized differences, see  Eq.~\eqref{rel_error_TH}, for (a) $E$ at $T=0.01$, (b) $S$ at $T=0.1$, and (c) $C_v$ at $T=0.2$ vs $l$ for the same expansions used in the main panels.}
    \label{fig:IS_contdis_all}
\end{figure}

In Fig.~\ref{fig:IS_contdis_all}, we plot the energy $E$, the entropy $S$, and the specific heat $C_v$ for the square lattice Ising model with a uniform disorder distribution. We show results for the highest two orders computed for the {\LL}, the square, and the rectangle expansions. For $E$ [see Fig.~\ref{fig:IS_contdis_all}(a)], the results for all expansions are indistinguishable from each other at temperatures down to $T=10^{-2}$, at which $E$ appears to saturate at the ground-state value (it is independent of $T$ at the lowest temperatures). For $S$ [see Fig.~\ref{fig:IS_contdis_all}(b)], small differences between the results for different expansions are seen for $T\lesssim 0.1$, but all the results converge towards $S=0$ as $T\rightarrow 0$ as expected. Similarly, for $C_v$ [see Fig.~\ref{fig:IS_contdis_all}(c)], small differences are seen below the maximum that occurs at $T_m\approx0.2$. The maximum of $C_v$ for the uniform disorder distribution occurs at a temperature lower than that [$T^{(b)}_m\approx0.4$] at which the maximum develops for bimodal disorder in Ref.~\cite{Tang_BiDisorder}. 

The insets in Fig.~\ref{fig:IS_contdis_all} show how the normalized difference for each observable, see Eq.~\eqref{rel_error_TH}, behaves with increasing the order $l$ of each expansion at a suitably chosen temperature (a temperature at which no differences are visible between the two orders of the NLCE in the main panels). The temperatures chosen are $T=0.01$ [inset in Fig.~\ref{fig:IS_contdis_all}(a)], $T=0.1$ [inset in Fig.~\ref{fig:IS_contdis_all}(b)], and $T=0.2$ [inset in Fig.~\ref{fig:IS_contdis_all}(c)]. In agreement with NLCE results obtained for translationally invariant models in earlier works~\cite{iyer_15,Abdelshafy}, one can see that for all observables and all NLCEs considered here the normalized differences decrease exponentially with $l$ before saturating due to the statistical errors at large $l$. As expected, given the faster convergence of NLCEs with increasing $T$, we find that the order $l$ at which such a saturation occurs decreases as the temperature increases (not shown).

The results in Fig.~\ref{fig:IS_contdis_all} make apparent that using a finite number of disorder realizations in the context of NLCEs with large building blocks, such as {\LL}s, squares, or rectangles allows us to obtain accurate results for the Ising model with a uniform disorder distribution down to $T\approx 10^{-2}$, at which the energy is nearly independent of the temperature, and the entropy and the specific heat are vanishingly small. In Appendix~\ref{sec:nonzero}, we report numerical results for the energy of the 2D square lattice Ising model with a continuous disorder distribution whose mean is nonzero. Those results are qualitatively similar to the ones reported in Fig.~\ref{fig:IS_contdis_all}(a), and they agree with Monte Carlo results for the same model and disorder distribution reported in Ref.~\cite{Mulanix_mltidis}.

\subsection{Heisenberg model}

\begin{figure}[!t]
    \includegraphics[width=.98\columnwidth]{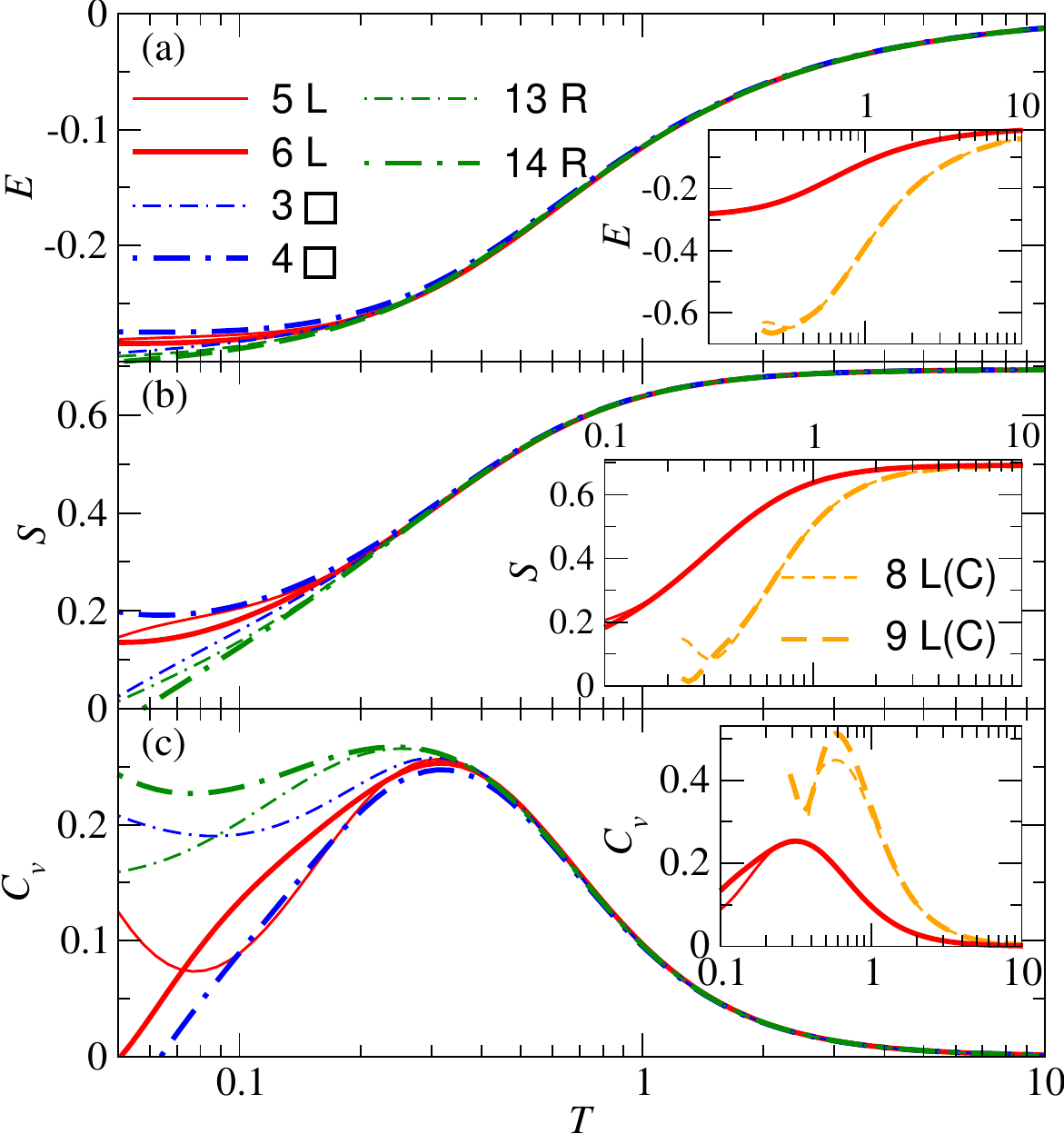}
    \vspace{-0.1cm}
    \caption{Thermodynamic properties of the square lattice Heisenberg model with a uniform disorder distribution. (a) Energy $E$, (b) entropy $S$, and (c) specific heat $C_v$ per site vs $T$ obtained using the {\LL} (\LL), the square ($\square$), and the rectangle (R) expansions. We report results for the highest two orders computed of each NLCE scheme. The insets contrast the results for the uniform disorder distribution to the corresponding ones in the clean case. For both sets of results the calculations were done using the {\LL} expansion.}
    \label{fig:AF_contdis_all}
\end{figure}

Next, we discuss our results for the most challenging model considered in this work. Namely, the square lattice Heisenberg model [see Eq.~\eqref{eq:HHeis}] with a uniform disorder distribution with zero mean [see Eq.~\eqref{eq:uniform}]. This model is frustrated and it is very challenging to study at low temperature using quantum Monte Carlo simulations because of the sign problem. In Fig.~\ref{fig:AF_contdis_all}, we show results for the highest two orders of the {\LL}, the square, and the rectangle expansions for the energy $E$, the entropy $S$, and the specific heat $C_v$. For the energy [see Fig.~\ref{fig:AF_contdis_all}(a)], the results for the three expansions are very close to each other down to $T \approx 0.2$, at which $E$ can be seen to begin to plateau to a temperature-independent value. For $S$ [see Fig.~\ref{fig:AF_contdis_all}(b)], all the results are also very close to one another down to $T\approx 0.2$. For $C_v$ [see Fig.~\ref{fig:AF_contdis_all}(c)], the results from different expansions depart from each other at temperatures $T\approx0.4$, below which a maximum appears to develop. Like for the entropy of the Heisenberg model with bimodal disorder in Fig.~\ref{fig:AF_S_bidis}, the {\LL} and square expansions are the closest ones for all observables in Fig.~\ref{fig:AF_contdis_all}.

In the insets in Fig.~\ref{fig:AF_contdis_all}, we contrast the {\LL} expansion results for the uniform disorder distribution to those obtained using the same expansion for the clean case. The effect of disorder in the square lattice can be seen to be qualitatively similar to that discussed in chains in the context of Fig.~\ref{fig:1D_AF}. Disorder increases the energies and entropies at all temperatures, as well as displaces the specific heat peak towards lower temperatures and reduces its height. In the presence of disorder, one can also see that the NLCE results for all observables converge at lower temperatures than in the clean case. This highlights a strength of NLCEs for systems with disorder, which in general have shorter correlations. For systems with disorder, NLCEs can provide accurate results to lower temperatures than for their clean counterparts.

\begin{figure}[!t]
    \includegraphics[width=.98\columnwidth]{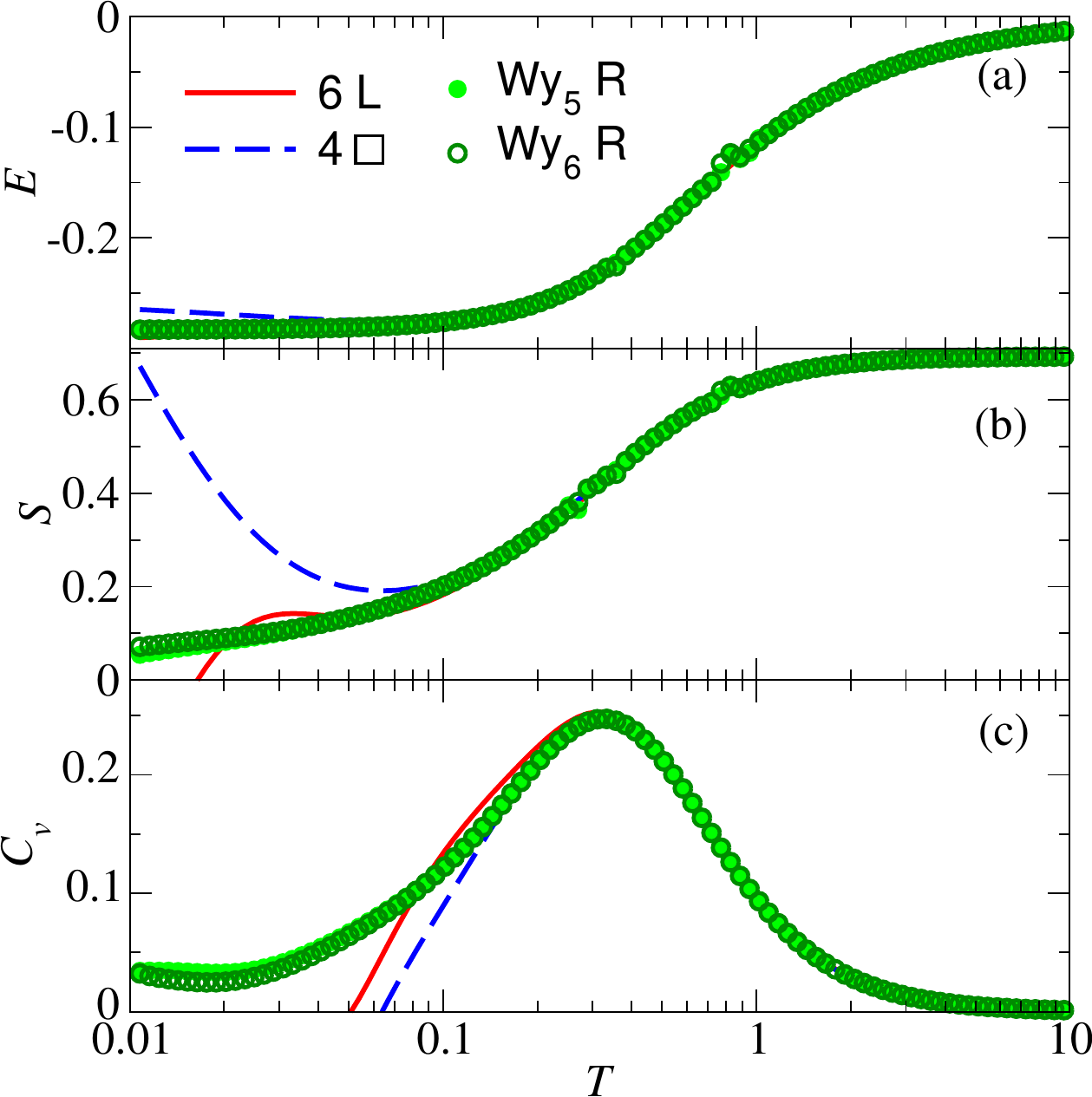}
    \vspace{-0.1cm}
    \caption{Thermodynamic properties of the square lattice Heisenberg model with a uniform disorder distribution. (a) Energy $E$, (b) entropy $S$, and (c) specific heat $C_v$ per site vs $T$. We show results for the highest order of the {\LL} and square expansions already shown in Fig.~\ref{fig:AF_contdis_all}, along with results of Euler's algorithm for the {\LL} (\LL) and the square ($\square$) expansions, and of Wynn's algorithm for the rectangle (R) expansion.}
    \label{fig:AF_contdis_resums}
\end{figure}

We applied Wynn and Euler resummation techniques to the NLCE results obtained for the square lattice Heisenberg model with a uniform disorder distribution. For all our observables within the rectangle expansion, Wynn's algorithm appears to extend the convergence to significantly lower temperatures than the direct sums. Unfortunately, since we have so few orders for the {\LL} and the square expansions, none of the resummation algorithms considered extended significantly the convergence of the corresponding direct sums. In Fig.~\ref{fig:AF_contdis_resums}, we compare the results of the highest order of the {\LL} and the square expansions (6 {\LL} and 4 $\square$, respectively) against those obtained for the highest two orders of Wynn's algorithm applied to the rectangle expansion results. 

For the energy [see Fig.~\ref{fig:AF_contdis_resums}(a)], the resummation results are very close to each other down to $T=10^{-2}$, and we see a clear plateau for temperatures between $T=10^{-2}$ and $T=10^{-1}$, so we expect the resummation results to be accurate all the way down to the ground-state energy. For the entropy [see Fig.~\ref{fig:AF_contdis_resums}(b)], the resummation results agree with each other down to $T\approx 0.04$, which is nearly an order of magnitude lower than that at which the direct sums agree with one another. Finally, given the behavior of the direct sums of the rectangle expansion for the specific heat in Fig.~\ref{fig:AF_contdis_all}(c), which appear to develop a maximum at lower temperatures than the {\LL} and square expansions, we find the most striking resummation results to be the ones for this observable. The resummation results for the rectangle expansion in Fig.~\ref{fig:AF_contdis_resums}(c) depart from those of their corresponding direct sums [shown in Fig.~\ref{fig:AF_contdis_all}(c)] below $T\approx 0.4$. Those resummation results are very close to the direct sums for the square expansion down to $T\approx0.2$, and very close to each other down to $T\approx 0.06$. The results for the direct sums and the resummations in Fig.~\ref{fig:AF_contdis_all}(c) suggest that a maximum occurs in the specific heat at $T_m\approx 0.3$. 

\section{Summary and discussion}\label{sec:summary}

We have shown that NLCEs based on sufficiently large building blocks allow one to obtain accurate low-temperature results for the thermodynamic properties of spin models with continuous disorder distributions in the square lattice. We used three NLCE schemes here, the restricted {\LL}, the square, and the rectangle expansions, and carried out the disorder averages directly on the NLCE clusters before computing their weights. We contrasted our results against those obtained using the site expansion, for which it was not possible to control the statistical errors because of the large number of clusters involved in low orders of the expansion. We advance that a similar approach can be used to study models with continuous disorder distributions in other lattice geometries, such as the triangular and kagome lattices, for which triangle-based expansions are readily available~\cite{rigol_bryant_07a, Abdelshafy}. 

We also showed that for the Ising model with an arbitrary disorder distribution in 1D, the linked cluster theorem provides an alternative way (to the traditional transfer matrix method) to obtain the exact analytical result for thermodynamic properties. For the Heisenberg model with a uniform disorder distribution in 1D, we provided evidence that NLCEs allow one to obtain the energy all the way down to the ground-state value, and the entropy and specific heat at temperatures that are about two orders of magnitude smaller than the value of $J$ used to set the width of the disorder distribution.

\acknowledgments

We acknowledge the support of the National Science Foundation Grants No.~PHY-2012145 and No.~PHY-2309146. The computations were done in the Institute for Computational and Data Sciences’ Roar supercomputer at Penn State.

\appendix

\section{Continuous disorder distributions\\ with nonzero mean}\label{sec:nonzero}

In this Appendix, we report additional results obtained using the {\LL}, the square, and the rectangle expansions for continuous disorder distributions that have a nonzero mean. We select those distributions, and their corresponding parameters, to be those for which site expansion and Monte Carlo results were reported in Ref.~\cite{Mulanix_mltidis}. The site expansion results in Ref.~\cite{Mulanix_mltidis} were obtained using multimodal disorder distributions. The results in this appendix allow one to contrast that approach to ours, with which we obtain results at lower temperatures.

\subsection{Ising model}

\begin{figure}[!t]
    \includegraphics[width=.98\columnwidth]{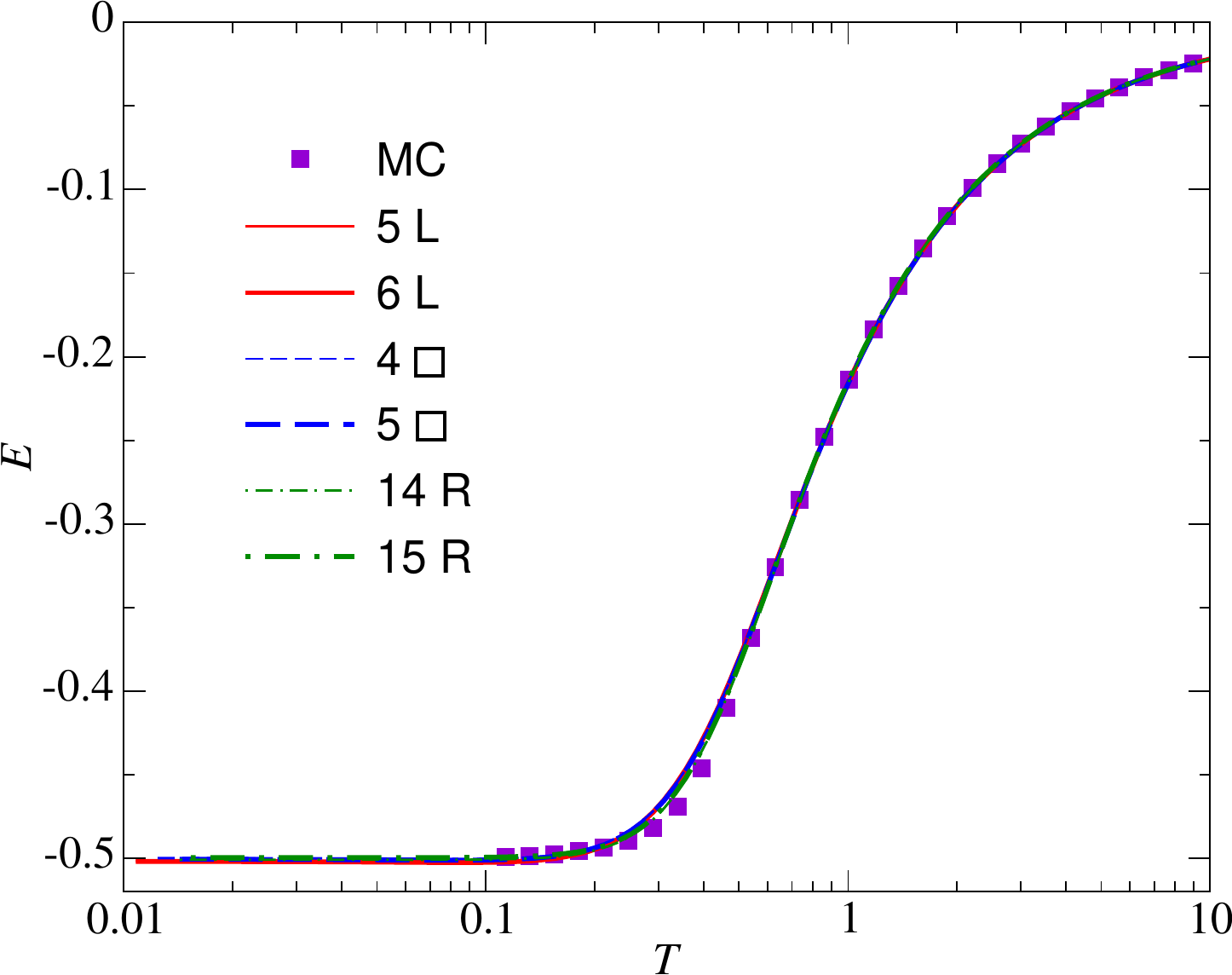}
    \vspace{-0.1cm}
    \caption{Energy $E$ per site vs $T$ for the square lattice Ising model with a uniform disorder distribution with nonzero mean (see text). We report results for the highest two orders of the {\LL} (\LL), square ($\square$), and rectangle (R) expansions along with the Monte Carlo (MC) results reported in Ref.~\cite{Mulanix_mltidis}.}
    \label{fig:IS_E_nonzeromean}
\end{figure}

For the Ising model, which being a classical model can be studied using Monte Carlo simulations, we consider the bond strengths in Eq.~\eqref{eq:HIsing} to be of the form $J_{\bf ij} = 1 + J\, R_{\bf ij}$, with $J = 1.5$ and $R_{\bf ij}$ drawn from the uniform distribution $[-1,1]$. In contrast to the case considered in the main text, this distribution exhibits more antiferromagnetic bonds than ferromagnetic ones. In Fig.~\ref{fig:IS_E_nonzeromean}, we show the energy per site vs the temperature from the highest two orders we computed of the {\LL}, the square, and the rectangle expansion. We contrast our results to those of Monte Carlo simulations from Ref.~\cite{Mulanix_mltidis}. In agreement with the multimodal NLCE results for the site expansion reported there (see Fig.~2 in Ref.~\cite{Mulanix_mltidis}), the results of our direct sums for the {\LL}, the square, and the rectangle expansions agree with the Monte Carlo ones at intermediate and high temperature ($T\gtrsim 0.5$). An advantage of the {\LL}, the square, and the rectangle expansions over the site expansion results in Ref.~\cite{Mulanix_mltidis} is that the former exhibit direct sums that allow us to compute the energy all the way down to the ground-state energy.

\subsection{Heisenberg model}

\begin{figure}[!h]
    \includegraphics[width=.98\columnwidth]{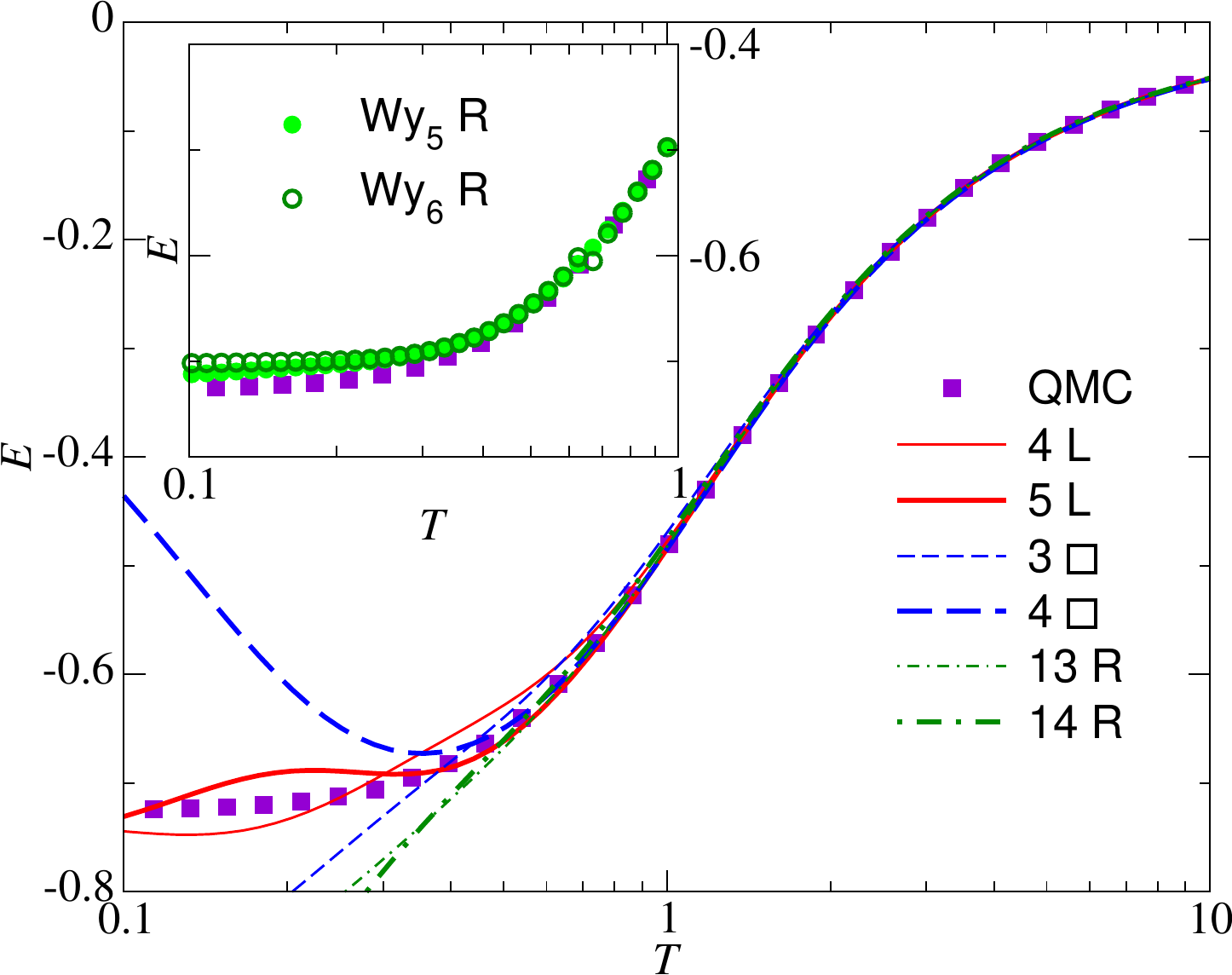}
    \vspace{-0.1cm}
    \caption{Energy $E$ per site vs $T$ for the square lattice Heisenberg model with a uniform disorder distribution with nonzero mean (see text). We report results for the highest two orders of the {\LL} (\LL), square ($\square$), and rectangle (R) expansions along with the Monte Carlo (QMC) results reported in Ref.~\cite{Mulanix_mltidis}. Inset: Results from Wynn's resummation of the rectangle expansion, Euler's resummations of the {\LL} and square expansions, and the Monte Carlo (QMC) results (same results and legends as in the main panel) reported in Ref.~\cite{Mulanix_mltidis}.}
    \label{fig:AF_E_nonzeromean}
\end{figure}

For the Heisenberg model, we consider the bond strengths in Eq.~\eqref{eq:HHeis} to be of the form $J_{\bf ij} = 1 + J\, R_{\bf ij}$, with $J = 1$ and $R_{\bf ij}$ drawn from the uniform distribution $[-1,1]$. For this selection of the disorder distribution, the model is not frustrated ($J_{\bf ij} \geq 0$, i.e., all the bonds remain antiferromagnetic) so accurate results can be obtained at all temperatures using quantum Monte Carlo (QMC) simulations. In Fig.~\ref{fig:AF_E_nonzeromean}, we show the energy per site vs the temperature from the highest two orders we computed of the {\LL}, the square, and the rectangle expansion. We contrast our results to those of QMC simulations (using the stochastic series expansions technique) from Ref.~\cite{Mulanix_mltidis}. Additionally, in the inset of Fig.~\ref{fig:AF_E_nonzeromean}, we show results obtained using the highest two orders of Wynn's resummation for the rectangle expansion.

Figure~\ref{fig:AF_E_nonzeromean} shows that the results of the direct sums for the highest order of the {\LL}, the square, and the rectangle expansion agree with each other and with the QMC results down to $T\approx 0.5$, which is about one half of the temperature at which the multimodal NLCE results for the site expansion agree with the QMC ones in Fig.~5 in Ref.~\cite{Mulanix_mltidis}. Like resummations in Ref.~\cite{Mulanix_mltidis}, in the inset in Fig.~\ref{fig:AF_E_nonzeromean} one can see that Wynn's resummations of the rectangle expansion extend the agreement of the NLCE results with the QMC ones to lower temperatures, $T\gtrsim 0.3$ in our case.\\

\newpage

\bibliographystyle{biblev1}
\bibliography{Reference}

\begin{thebibliography}{10}
\expandafter\ifx\csname url\endcsname\relax
  \def\url#1{{\tt #1}}\fi
\expandafter\ifx\csname urlprefix\endcsname\relax\def\urlprefix{URL }\fi
\expandafter\ifx\csname bibinfo\endcsname\relax\def\bibinfo#1#2{#2}\fi
\expandafter\ifx\csname eprint\endcsname\relax\def\eprint#1{\url{#1}}\fi

\bibitem{Anderson}
\bibinfo{author}{P.~W. Anderson}, \bibinfo{title}{Absence of diffusion in
  certain random lattices},
  \bibinfo{journal}{\href{http://dx.doi.org/10.1103/PhysRev.109.1492}{Phys.
  Rev.}} \href{http://dx.doi.org/10.1103/PhysRev.109.1492}{{\bf
  \bibinfo{volume}{109}}, \bibinfo{pages}{1492}}
  (\href{http://dx.doi.org/10.1103/PhysRev.109.1492}{\bibinfo{year}{1958}}).

\bibitem{Edwards_1975}
\bibinfo{author}{S.~F. Edwards} and \bibinfo{author}{P.~W. Anderson},
  \bibinfo{title}{Theory of spin glasses},
  \bibinfo{journal}{\href{http://dx.doi.org/10.1088/0305-4608/5/5/017}{Journal
  of Physics F: Metal Physics}}
  \href{http://dx.doi.org/10.1088/0305-4608/5/5/017}{{\bf \bibinfo{volume}{5}},
  \bibinfo{pages}{965}}
  (\href{http://dx.doi.org/10.1088/0305-4608/5/5/017}{\bibinfo{year}{1975}}).

\bibitem{Binder_spinglass}
\bibinfo{author}{K.~Binder} and \bibinfo{author}{A.~P. Young},
  \bibinfo{title}{Spin glasses: Experimental facts, theoretical concepts, and
  open questions},
  \bibinfo{journal}{\href{http://dx.doi.org/10.1103/RevModPhys.58.801}{Rev.
  Mod. Phys.}} \href{http://dx.doi.org/10.1103/RevModPhys.58.801}{{\bf
  \bibinfo{volume}{58}}, \bibinfo{pages}{801}}
  (\href{http://dx.doi.org/10.1103/RevModPhys.58.801}{\bibinfo{year}{1986}}).

\bibitem{Weissman_spinglass}
\bibinfo{author}{M.~B. Weissman}, \bibinfo{title}{What is a spin glass? a
  glimpse via mesoscopic noise},
  \bibinfo{journal}{\href{http://dx.doi.org/10.1103/RevModPhys.65.829}{Rev.
  Mod. Phys.}} \href{http://dx.doi.org/10.1103/RevModPhys.65.829}{{\bf
  \bibinfo{volume}{65}}, \bibinfo{pages}{829}}
  (\href{http://dx.doi.org/10.1103/RevModPhys.65.829}{\bibinfo{year}{1993}}).

\bibitem{loh_90}
\bibinfo{author}{E.~Y. Loh}, \bibinfo{author}{J.~E. Gubernatis},
  \bibinfo{author}{R.~T. Scalettar}, \bibinfo{author}{S.~R. White},
  \bibinfo{author}{D.~J. Scalapino}, and \bibinfo{author}{R.~L. Sugar},
  \bibinfo{title}{Sign problem in the numerical simulation of many-electron
  systems},
  \bibinfo{journal}{\href{http://dx.doi.org/10.1103/PhysRevB.41.9301}{Phys.
  Rev. B}} \href{http://dx.doi.org/10.1103/PhysRevB.41.9301}{{\bf
  \bibinfo{volume}{41}}, \bibinfo{pages}{9301}}
  (\href{http://dx.doi.org/10.1103/PhysRevB.41.9301}{\bibinfo{year}{1990}}).

\bibitem{henelius_00}
\bibinfo{author}{P.~Henelius} and \bibinfo{author}{A.~W. Sandvik},
  \bibinfo{title}{Sign problem in {M}onte {C}arlo simulations of frustrated
  quantum spin systems},
  \bibinfo{journal}{\href{http://dx.doi.org/10.1103/PhysRevB.62.1102}{Phys.
  Rev. B}} \href{http://dx.doi.org/10.1103/PhysRevB.62.1102}{{\bf
  \bibinfo{volume}{62}}, \bibinfo{pages}{1102}}
  (\href{http://dx.doi.org/10.1103/PhysRevB.62.1102}{\bibinfo{year}{2000}}).

\bibitem{troyer_05}
\bibinfo{author}{M.~Troyer} and \bibinfo{author}{U.-J. Wiese},
  \bibinfo{title}{Computational complexity and fundamental limitations to
  fermionic quantum {M}onte {C}arlo simulations},
  \bibinfo{journal}{\href{http://dx.doi.org/10.1103/PhysRevLett.94.170201}{Phys.
  Rev. Lett.}} \href{http://dx.doi.org/10.1103/PhysRevLett.94.170201}{{\bf
  \bibinfo{volume}{94}}, \bibinfo{pages}{170201}}
  (\href{http://dx.doi.org/10.1103/PhysRevLett.94.170201}{\bibinfo{year}{2005}}).

\bibitem{Tang_dynamicdis}
\bibinfo{author}{B.~Tang}, \bibinfo{author}{D.~Iyer}, and
  \bibinfo{author}{M.~Rigol}, \bibinfo{title}{Quantum quenches and many-body
  localization in the thermodynamic limit},
  \bibinfo{journal}{\href{http://dx.doi.org/10.1103/PhysRevB.91.161109}{Phys.
  Rev. B}} \href{http://dx.doi.org/10.1103/PhysRevB.91.161109}{{\bf
  \bibinfo{volume}{91}}, \bibinfo{pages}{161109}}
  (\href{http://dx.doi.org/10.1103/PhysRevB.91.161109}{\bibinfo{year}{2015}}).

\bibitem{Tang_BiDisorder}
\bibinfo{author}{B.~Tang}, \bibinfo{author}{D.~Iyer}, and
  \bibinfo{author}{M.~Rigol}, \bibinfo{title}{Thermodynamics of two-dimensional
  spin models with bimodal random-bond disorder},
  \bibinfo{journal}{\href{http://dx.doi.org/10.1103/PhysRevB.91.174413}{Phys.
  Rev. B}} \href{http://dx.doi.org/10.1103/PhysRevB.91.174413}{{\bf
  \bibinfo{volume}{91}}, \bibinfo{pages}{174413}}
  (\href{http://dx.doi.org/10.1103/PhysRevB.91.174413}{\bibinfo{year}{2015}}).

\bibitem{Mulanix_mltidis}
\bibinfo{author}{M.~D. Mulanix}, \bibinfo{author}{D.~Almada}, and
  \bibinfo{author}{E.~Khatami}, \bibinfo{title}{Numerical linked-cluster
  expansions for disordered lattice models},
  \bibinfo{journal}{\href{http://dx.doi.org/10.1103/PhysRevB.99.205113}{Phys.
  Rev. B}} \href{http://dx.doi.org/10.1103/PhysRevB.99.205113}{{\bf
  \bibinfo{volume}{99}}, \bibinfo{pages}{205113}}
  (\href{http://dx.doi.org/10.1103/PhysRevB.99.205113}{\bibinfo{year}{2019}}).

\bibitem{park_khatami_21}
\bibinfo{author}{J.~Park} and \bibinfo{author}{E.~Khatami},
  \bibinfo{title}{Thermodynamics of the disordered hubbard model studied via
  numerical linked-cluster expansions},
  \bibinfo{journal}{\href{http://dx.doi.org/10.1103/PhysRevB.104.165102}{Phys.
  Rev. B}} \href{http://dx.doi.org/10.1103/PhysRevB.104.165102}{{\bf
  \bibinfo{volume}{104}}, \bibinfo{pages}{165102}}
  (\href{http://dx.doi.org/10.1103/PhysRevB.104.165102}{\bibinfo{year}{2021}}).

\bibitem{rigol_bryant_06}
\bibinfo{author}{M.~Rigol}, \bibinfo{author}{T.~Bryant}, and
  \bibinfo{author}{R.~R.~P. Singh}, \bibinfo{title}{Numerical linked-cluster
  approach to quantum lattice models},
  \bibinfo{journal}{\href{http://dx.doi.org/10.1103/PhysRevLett.97.187202}{Phys.
  Rev. Lett.}} \href{http://dx.doi.org/10.1103/PhysRevLett.97.187202}{{\bf
  \bibinfo{volume}{97}}, \bibinfo{pages}{187202}}
  (\href{http://dx.doi.org/10.1103/PhysRevLett.97.187202}{\bibinfo{year}{2006}}).

\bibitem{rigol_bryant_07a}
\bibinfo{author}{M.~Rigol}, \bibinfo{author}{T.~Bryant}, and
  \bibinfo{author}{R.~R.~P. Singh}, \bibinfo{title}{Numerical linked-cluster
  algorithms. {I. S}pin systems on square, triangular, and kagom\'e lattices},
  \bibinfo{journal}{\href{http://dx.doi.org/10.1103/PhysRevE.75.061118}{Phys.
  Rev. E}} \href{http://dx.doi.org/10.1103/PhysRevE.75.061118}{{\bf
  \bibinfo{volume}{75}}, \bibinfo{pages}{061118}}
  (\href{http://dx.doi.org/10.1103/PhysRevE.75.061118}{\bibinfo{year}{2007}}).

\bibitem{rigol_bryant_07b}
\bibinfo{author}{M.~Rigol}, \bibinfo{author}{T.~Bryant}, and
  \bibinfo{author}{R.~R.~P. Singh}, \bibinfo{title}{Numerical linked-cluster
  algorithms. {II. $t\text{\ensuremath{-}}J$} models on the square lattice},
  \bibinfo{journal}{\href{http://dx.doi.org/10.1103/PhysRevE.75.061119}{Phys.
  Rev. E}} \href{http://dx.doi.org/10.1103/PhysRevE.75.061119}{{\bf
  \bibinfo{volume}{75}}, \bibinfo{pages}{061119}}
  (\href{http://dx.doi.org/10.1103/PhysRevE.75.061119}{\bibinfo{year}{2007}}).

\bibitem{Abdelshafy}
\bibinfo{author}{M.~Abdelshafy} and \bibinfo{author}{M.~Rigol},
  \bibinfo{title}{L-based numerical linked cluster expansion for square lattice
  models},
  \bibinfo{journal}{\href{http://dx.doi.org/10.1103/PhysRevE.108.034126}{Phys.
  Rev. E}} \href{http://dx.doi.org/10.1103/PhysRevE.108.034126}{{\bf
  \bibinfo{volume}{108}}, \bibinfo{pages}{034126}}
  (\href{http://dx.doi.org/10.1103/PhysRevE.108.034126}{\bibinfo{year}{2023}}).

\bibitem{devakul_15}
\bibinfo{author}{T.~Devakul} and \bibinfo{author}{R.~R.~P. Singh},
  \bibinfo{title}{Early breakdown of area-law entanglement at the many-body
  delocalization transition},
  \bibinfo{journal}{\href{http://dx.doi.org/10.1103/PhysRevLett.115.187201}{Phys.
  Rev. Lett.}} \href{http://dx.doi.org/10.1103/PhysRevLett.115.187201}{{\bf
  \bibinfo{volume}{115}}, \bibinfo{pages}{187201}}
  (\href{http://dx.doi.org/10.1103/PhysRevLett.115.187201}{\bibinfo{year}{2015}}).

\bibitem{Hazzard_dis}
\bibinfo{author}{J.~Gan} and \bibinfo{author}{K.~R.~A. Hazzard},
  \bibinfo{title}{Numerical linked cluster expansions for inhomogeneous
  systems},
  \bibinfo{journal}{\href{http://dx.doi.org/10.1103/PhysRevA.102.013318}{Phys.
  Rev. A}} \href{http://dx.doi.org/10.1103/PhysRevA.102.013318}{{\bf
  \bibinfo{volume}{102}}, \bibinfo{pages}{013318}}
  (\href{http://dx.doi.org/10.1103/PhysRevA.102.013318}{\bibinfo{year}{2020}}).

\bibitem{Pardini_19}
\bibinfo{author}{T.~Pardini}, \bibinfo{author}{A.~Menon},
  \bibinfo{author}{S.~P. Hau-Riege}, and \bibinfo{author}{R.~R.~P. Singh},
  \bibinfo{title}{Local entanglement and confinement transitions in the random
  transverse-field {I}sing model on the pyrochlore lattice},
  \bibinfo{journal}{\href{http://dx.doi.org/10.1103/PhysRevB.100.144437}{Phys.
  Rev. B}} \href{http://dx.doi.org/10.1103/PhysRevB.100.144437}{{\bf
  \bibinfo{volume}{100}}, \bibinfo{pages}{144437}}
  (\href{http://dx.doi.org/10.1103/PhysRevB.100.144437}{\bibinfo{year}{2019}}).

\bibitem{kallin_13}
\bibinfo{author}{A.~B. Kallin}, \bibinfo{author}{K.~Hyatt},
  \bibinfo{author}{R.~R.~P. Singh}, and \bibinfo{author}{R.~G. Melko},
  \bibinfo{title}{Entanglement at a two-dimensional quantum critical point: {A}
  numerical linked-cluster expansion study},
  \bibinfo{journal}{\href{http://dx.doi.org/10.1103/PhysRevLett.110.135702}{Phys.
  Rev. Lett.}} \href{http://dx.doi.org/10.1103/PhysRevLett.110.135702}{{\bf
  \bibinfo{volume}{110}}, \bibinfo{pages}{135702}}
  (\href{http://dx.doi.org/10.1103/PhysRevLett.110.135702}{\bibinfo{year}{2013}}).

\bibitem{Seing_Rect}
\bibinfo{author}{J.~Richter}, \bibinfo{author}{T.~Heitmann}, and
  \bibinfo{author}{R.~Steinigeweg}, \bibinfo{title}{Quantum quench dynamics in
  the transverse-field {I}sing model: {A} numerical expansion in linked
  rectangular clusters},
  \bibinfo{journal}{\href{http://dx.doi.org/10.21468/SciPostPhys.9.3.031}{SciPost
  Phys.}} \href{http://dx.doi.org/10.21468/SciPostPhys.9.3.031}{{\bf
  \bibinfo{volume}{9}}, \bibinfo{pages}{031}}
  (\href{http://dx.doi.org/10.21468/SciPostPhys.9.3.031}{\bibinfo{year}{2020}}).

\bibitem{iyer_15}
\bibinfo{author}{D.~Iyer}, \bibinfo{author}{M.~Srednicki}, and
  \bibinfo{author}{M.~Rigol}, \bibinfo{title}{Optimization of finite-size
  errors in finite-temperature calculations of unordered phases},
  \bibinfo{journal}{\href{http://dx.doi.org/10.1103/PhysRevE.91.062142}{Phys.
  Rev. E}} \href{http://dx.doi.org/10.1103/PhysRevE.91.062142}{{\bf
  \bibinfo{volume}{91}}, \bibinfo{pages}{062142}}
  (\href{http://dx.doi.org/10.1103/PhysRevE.91.062142}{\bibinfo{year}{2015}}).

\bibitem{Tang_2013}
\bibinfo{author}{B.~Tang}, \bibinfo{author}{E.~Khatami}, and
  \bibinfo{author}{M.~Rigol}, \bibinfo{title}{A short introduction to numerical
  linked-cluster expansions},
  \bibinfo{journal}{\href{http://dx.doi.org/10.1016/j.cpc.2012.10.008}{Computer
  Physics Communications}}
  \href{http://dx.doi.org/10.1016/j.cpc.2012.10.008}{{\bf
  \bibinfo{volume}{184}}, \bibinfo{pages}{557}}
  (\href{http://dx.doi.org/10.1016/j.cpc.2012.10.008}{\bibinfo{year}{2013}}).

\bibitem{Kowalski_1975}
\bibinfo{author}{J.~M. Kowalski} and \bibinfo{author}{A.~Pekalski},
  \bibinfo{title}{One-dimensional {I}sing model with random bonds},
  \bibinfo{journal}{\href{http://dx.doi.org/10.1088/0022-3719/8/13/018}{J Phys.
  C Solid State Phys.}}
  \href{http://dx.doi.org/10.1088/0022-3719/8/13/018}{{\bf
  \bibinfo{volume}{8}}, \bibinfo{pages}{2085}}
  (\href{http://dx.doi.org/10.1088/0022-3719/8/13/018}{\bibinfo{year}{1975}}).

\bibitem{Ma_79}
\bibinfo{author}{S.-k. Ma}, \bibinfo{author}{C.~Dasgupta}, and
  \bibinfo{author}{C.-k. Hu}, \bibinfo{title}{Random antiferromagnetic chain},
  \bibinfo{journal}{\href{http://dx.doi.org/10.1103/PhysRevLett.43.1434}{Phys.
  Rev. Lett.}} \href{http://dx.doi.org/10.1103/PhysRevLett.43.1434}{{\bf
  \bibinfo{volume}{43}}, \bibinfo{pages}{1434}}
  (\href{http://dx.doi.org/10.1103/PhysRevLett.43.1434}{\bibinfo{year}{1979}}).

\bibitem{Dasgupta_Ma_80}
\bibinfo{author}{C.~Dasgupta} and \bibinfo{author}{S.-k. Ma},
  \bibinfo{title}{Low-temperature properties of the random {H}eisenberg
  antiferromagnetic chain},
  \bibinfo{journal}{\href{http://dx.doi.org/10.1103/PhysRevB.22.1305}{Phys.
  Rev. B}} \href{http://dx.doi.org/10.1103/PhysRevB.22.1305}{{\bf
  \bibinfo{volume}{22}}, \bibinfo{pages}{1305}}
  (\href{http://dx.doi.org/10.1103/PhysRevB.22.1305}{\bibinfo{year}{1980}}).

\bibitem{Giamarchi_03}
\bibinfo{author}{T.~Giamarchi}, {\em \bibinfo{title}{{Quantum Physics in One
  Dimension}}\/} (\bibinfo{publisher}{Oxford University Press},
  \bibinfo{year}{2003}).

\end{thebibliography}

\end{document}